\documentclass[aps,pre,preprint,showkeys,nofootinbib,byrevtex]{revtex4}

\usepackage{amsmath,amssymb,amsfonts,graphicx,framed}
\usepackage{array}

\begin{document}

\title{Harmonic Standing-Wave Excitations of Simply-Supported Thick-Walled Hollow Elastic Circular Cylinders: \\ Exact 3D Linear Elastodynamic Response}

\author{Jamal Sakhr}
\author{Blaine A. Chronik}
\affiliation{Department of Physics and Astronomy, The University of Western Ontario, London, Ontario, Canada N6A 3K7}

\date{\today} 

\begin{abstract}
The forced-vibration response of a simply-supported isotropic thick-walled hollow elastic circular cylinder subjected to two-dimensional harmonic standing-wave excitations on its curved surfaces is studied within the framework of linear elastodynamics. Exact semi-analytical solutions for the steady-state displacement field of the cylinder are constructed using recently-published parametric solutions to the Navier-Lam\'{e} equation. Formal application of the standing-wave boundary conditions generates three parameter-dependent $6\times6$ linear systems, each of which can be numerically solved in order to determine the parametric response of the cylinder's displacement field under various conditions. The method of solution is direct and demonstrates a general approach that can be applied to solve many other elastodynamic forced-response problems involving isotropic elastic cylinders. As an application, and considering several examples, the obtained solution is used to compute the steady-state frequency response in a few specific low-order excitation cases. In each case, the solution generates a series of resonances that are in exact correspondence with a unique subset of the natural frequencies of the simply-supported cylinder. The considered problem is of general theoretical interest in structural mechanics and acoustics and more practically serves as a benchmark forced-vibration problem involving a thick-walled hollow elastic cylinder. 
\end{abstract} 


\maketitle

\section{Introduction}\label{intro}

The vibration of an isotropic thick-walled hollow elastic circular cylinder is one of the classical applied problems of elastodynamics and has been of longstanding general interest to applied mathematicians, acousticians, engineers, and physicists \cite{Soldatos94,Qatu02,VTCS10,LeissaQatu11}. There is a vast literature on the \emph{free} vibration of finite-length isotropic hollow elastic circular cylinders, and while much of the fundamental work on the subject was carried out prior to the 1980s, there has, over the last three decades, been a steady stream of publications devoted to developing and testing various methodological approaches to obtaining natural frequencies and mode shapes under a variety of end conditions (see, for example, Refs.~\cite{Hutch86,Soldatos90,Leissa97,Loy99,Buchanan02,Zhou03,Mofak06,MFZK08,Khalili12,Ye14} and references therein). The literature on the \emph{forced} vibration of hollow elastic cylinders (finite-length, isotropic, circular, or otherwise) is much smaller in comparison. Forced-vibration analyses of thick-walled elastic cylinders based on the exact three-dimensional (3D) theory of linear elasticity, in particular, remain scarce. In the context of forced-vibration analyses based on the Navier-Lam\'{e} equation of motion, the most notable and general work is due to Ebenezer and co-workers \cite{Ebenezer15} (ERP), who devised an exact series method to determine the steady-state vibration response of a finite-length isotropic hollow elastic circular cylinder subjected to arbitrary \emph{axisymmetric} excitations on its surfaces. 
%In recent years, there has been a lot of research on the free- and forced-vibration of \emph{thin} cylindrical shells with complex end conditions (see, for example, Refs~\cite{BBB} and references therein). Unfortunately, many of the shell theories and/or mathematical formulations employed in this domain of research are inappropriate or inaccurate when applied to thick-walled cylinders, and further exact analyses based on 3D linear elasticity are needed. 

As pointed out in many reviews (see, for example, Ref.~\cite{VTCS10}), many fundamental forced-vibration problems involving hollow elastic cylinders have not yet been studied or solved using the exact 3D theory of linear elasticity. A useful and analytically tractable model problem that has been surprisingly overlooked is the steady-state vibration response problem for a simply-supported isotropic thick-walled hollow elastic circular cylinder subjected to arbitrary excitations on its curved surfaces. For a simply-supported cylinder, arbitrary \emph{asymmetric} excitations on the curved surfaces can be naturally expressed as superpositions of two-dimensional (2D) harmonic standing waves in the circumferential and axial directions. Thus, the precursor is to consider individual 2D harmonic standing-wave excitations on the curved surfaces. This latter problem is theoretically significant in its own right since it is one of the few model problems for which the effects of \emph{individual} harmonic excitations on the curved surfaces of the cylinder can be isolated and studied without having to incorporate non-trivial corrections in order to simultaneously satisfy the end conditions. In general, the exact nature of these effects will be obscured by other excitations needed to generate the desired end conditions.\footnote{Incidentally, a corollary of ERP's work \cite{Ebenezer15} is that arbitrary \emph{axisymmetric} excitations on the curved surfaces of a finite-length hollow elastic cylinder can be mathematically expressed as (finite or infinite) superpositions of one-dimensional harmonic standing waves in the axial direction. Curiously, the effect of the \emph{individual} harmonics on the steady-state vibration response was not considered in Ref.~\cite{Ebenezer15}.} 

The free-vibration analog of the proposed problem, that is, the free-vibration problem for a simply-supported isotropic (thick-walled) hollow elastic circular cylinder, is an important benchmark problem in many numerical free-vibration studies (see, for example, Ref.~\cite{Khalili12} and references therein). Explicit analytical formulations and mathematical analyses of this problem are however not easy to find in the literature. Weingarten and Reismann \cite{Wein74} applied the method of eigenfunction expansions to this problem and obtained an implicit solution in 1974. The free-vibration analog can also be extracted as a special case of the largely overlooked work of Prasad and Jain \cite{PJ60s}, who, already in the mid 1960s, considered the problem of \emph{free} harmonic standing waves in a simply-supported transversely isotropic hollow elastic circular cylinder. In the two aforementioned works, the authors did not actually solve their obtained frequency equations, and thus, did not explicitly obtain any natural frequencies or mode shapes. These have however since been obtained both indirectly\footnote{It is a well-known and often-cited fact that the characteristics of free harmonic \emph{standing} waves in a finite-length simply-supported hollow elastic cylinder are formally equivalent to those of free harmonic \emph{traveling} waves in a corresponding cylinder of infinite length, the latter of which are well-studied (see, for example, Ref.~\cite{Arm69}). Numerical solutions to the free standing-wave problem are thus readily available as a corollary (see Ref.~\cite{Soldatos94} for further discussions).} and through the use of specialized methods tailored to solving the general free-vibration problem for an isotropic hollow elastic circular cylinder \cite{Soldatos90,Loy99,Khalili12,Ye14}. 

Although hitherto unstudied, the problem of interest in this paper is not without precedent. Hamidzadeh \textit{et al.} (see Ref.~\cite{VTCS10} and references therein) considered the problem of determining the resonant frequencies of an infinitely-long isotropic thick-walled elastic circular cylinder subjected to harmonic boundary stresses. Using a ``frequency sweeping'' procedure (see Ref.~\cite{VTCS10} for details), they calculated the resonant frequencies and subsequently compared them to the natural frequencies given in Ref.~\cite{Arm69}; dissimilarities were observed for short cylinders and fundamental resonant modes. The boundary-value problem considered by Hamidzadeh \textit{et al.} is the infinite-cylinder analog of the boundary-value problem that we seek to study in this paper. 

Before entering into details, it is useful to give a brief overview of the paper. In Section~\ref{EGFULL}, we define the boundary-value problem of interest. In Sections~\ref{GenFormDispField}-\ref{solntoBVPKeq0}, we construct, exploiting certain known solutions to the Navier-Lam\'{e} equation \cite{usB1} (Section~\ref{SOLNSgen}), an exact semi-analytical 3D elastodynamic solution to the problem. The method of solution is direct and demonstrates a general approach that can be applied to solve other similar forced-vibration problems involving elastic cylinders. The solution itself, although exact and given in closed form, involves six constants whose values are not given in explicit analytical form. It is in this sense that the obtained solution is a ``semi-analytical'' solution. Some numerical examples are subsequently given in Sections~\ref{numerica}-\ref{NumEG2}, wherein the obtained solution is used to study the steady-state frequency response in some example excitation cases. In each case, consistency with published natural frequency data is observed. More detailed conclusions based on our numerical investigations are summarized in Section~\ref{conclus}. 

\section{Mathematical Definition of the Problem}\label{EGFULL}

Consider a simply-supported isotropic hollow elastic circular cylinder of length $L$ and inner and outer radii $R_1$ and $R_2$, respectively. The geometrical parameters $\{L,R_1,R_2\}$ are all finite, but otherwise arbitrary. Suppose the cylinder is subjected to time-harmonic stresses on its curved surfaces and furthermore that these stresses are spatially non-uniform and are such that the circumferential and longitudinal variations are also harmonic. In this paper, we shall work in the circular cylindrical coordinate system wherein all physical quantities depend on the spatial coordinates $(r,\theta,z)$, which denote the radial, circumferential, and longitudinal coordinates, respectively, and on the time $t$. Using this notation, the boundary stresses on the curved surfaces are: 
\begin{subequations}\label{BCsCRVD2}
\begin{eqnarray}\label{strssrr2}
\left\{\begin{array}{c}\sigma_{rr}(R_1,\theta,z,t) \\ \sigma_{rr}(R_2,\theta,z,t) \end{array} \right\}=\left \{\begin{array}{c} \mathcal{A} \\ \mathcal{D} \end{array} \right\} \cos(m\theta)\sin\left({k\pi\over L}z\right)\sin(\omega t),
\end{eqnarray}
\begin{eqnarray}\label{strssrt2}
\left\{\begin{array}{c} \sigma_{r\theta}(R_1,\theta,z,t) \\ \sigma_{r\theta}(R_2,\theta,z,t) \end{array}\right\}=\left \{\begin{array}{c} \mathcal{B} \\ \mathcal{E} \end{array} \right\}\sin(m\theta)\sin\left({k\pi\over L}z\right)\sin(\omega t), 
\end{eqnarray}
\begin{eqnarray}\label{strssrz2}
\left\{\begin{array}{c} \sigma_{rz}(R_1,\theta,z,t) \\ \sigma_{rz}(R_2,\theta,z,t)\end{array}\right\}=\left \{\begin{array}{c} \mathcal{C} \\ \mathcal{F} \end{array} \right\}\cos(m\theta)\cos\left({k\pi\over L}z\right)\sin(\omega t), 
\end{eqnarray}
\end{subequations}
where $\sigma_{rr}(r,\theta,z,t)$ is a normal component of stress, $\sigma_{r\theta}(r,\theta,z,t)$ and $\sigma_{rz}(r,\theta,z,t)$ are shear components of stress, $\{\mathcal{A},\mathcal{B},\mathcal{C},\mathcal{D},\mathcal{E},\mathcal{F}\}$ are prescribed constant stresses (each having units of pressure), $k$ and $m$ are prescribed non-negative integers, and $\omega>0$ is the prescribed angular frequency of excitation. Since the harmonic temporal variation is separable from the harmonic spatial variations in each of the excitations (\ref{strssrr2})-(\ref{strssrz2}), these represent harmonic standing-wave excitations. 

The specific problem of interest here is to determine the elastodynamic response of the cylinder when it is subjected to the non-uniform distribution of stress (\ref{BCsCRVD2}) on its curved surfaces. In other words, we seek to determine the (frequency-dependent) displacement field at all points of the cylinder. The governing equation of motion for the displacement is the Navier-Lam\'{e} (NL) equation, which can be written in vector form as \cite{Elastobuch}:
\begin{equation}\label{NLE}
(\lambda+2\mu)\nabla(\nabla\cdot\mathbf{u})-\mu\nabla\times(\nabla\times\mathbf{u})+\mathbf{b}=\rho{\partial^2\mathbf{u}\over\partial t^2},
\end{equation}
where $\mathbf{u}\equiv\mathbf{u}(r,\theta,z,t)$ is the displacement field, $\lambda>0$ and $\mu>0$ are the first and second Lam\'{e} constants, respectively\footnote{Note that the first Lam\'{e} constant $\lambda$ need not be positive, but we have assumed it to be so for the purposes of this paper.}, and $\rho>0$ is the (constant) density of the cylinder. Since there are only surface forces acting on the cylinder, the local body force is zero (i.e., $\mathbf{b}=\mathbf{0}$). The radial, circumferential, and longitudinal components of $\mathbf{u}$ shall here be denoted by $u_r(r,\theta,z,t),u_\theta(r,\theta,z,t)$, and $u_z(r,\theta,z,t)$, respectively. 

Although we have stated that the cylinder is simply supported, we have not yet specified the boundary conditions at the flat ends of the cylinder, which are situated at $z=0$ and $z=L$. The classical simply-supported (SS) boundary conditions for the stress and displacement at the flat ends of the cylinder are: 
%\footnote{Although perhaps impracticable, SS end conditions are mathematically permissible and are widely-used, especially in free-vibration analyses of isotropic hollow elastic circular cylinders (see, for example, Refs.~\cite{Soldatos90,Loy99,MFZK08,Khalili12,Ye14}).}:
\begin{subequations}\label{SSBCs}
\begin{equation}\label{SSBCs1}
u_r(r,\theta,0,t)=u_r(r,\theta,L,t)=0,
\end{equation}
\begin{equation}\label{SSBCs2}
u_\theta(r,\theta,0,t)=u_\theta(r,\theta,L,t)=0,
\end{equation}
\begin{equation}\label{SSBCs3}
\sigma_{zz}(r,\theta,0,t)=\sigma_{zz}(r,\theta,L,t)=0,
\end{equation}
\end{subequations}
where $\sigma_{zz}(r,\theta,z,t)$ is the normal component of stress along the axis of the cylinder.  

Conditions (\ref{strssrr2})-(\ref{strssrz2}) must be satisfied for all $\theta\in[0,2\pi]$, $z\in(0,L)$, and arbitrary $t$.  Conditions (\ref{SSBCs1})-(\ref{SSBCs3}) must be satisfied for all $r\in[R_1,R_2]$, $\theta\in[0,2\pi]$, and arbitrary $t$. Note that we have not given any information about the displacement field and its time derivatives at some initial time $t=t_0$, and thus the problem as defined is not an initial-boundary-value problem. 
Note that, for forced motion, \emph{at least} one of $\{\mathcal{A},\mathcal{B},\mathcal{C},\mathcal{D},\mathcal{E},\mathcal{F}\}$ must be non-zero when $m\neq0$. If $m=0$, then \emph{at least} one of $\{\mathcal{A},\mathcal{C},\mathcal{D},\mathcal{F}\}$ is required to be non-zero in boundary conditions (\ref{BCsCRVD2}). 

For future reference, we cite here the cylindrical stress-displacement relations from the linear theory of elasticity \cite{VTCS10}:
\begin{subequations}\label{strssdispCYL}
\begin{equation}\label{stssstrnCYL1}
\sigma_{rr}=(\lambda+2\mu){\partial u_r\over\partial r}+{\lambda\over r}\left({\partial u_\theta\over\partial \theta}+u_r\right)+\lambda{\partial u_z\over\partial z},
\end{equation}
\begin{equation}\label{stssstrnCYL2}
\sigma_{\theta\theta}=\lambda{\partial u_r\over\partial r}+{(\lambda+2\mu)\over r}\left({\partial u_\theta\over\partial \theta}+u_r\right)+\lambda{\partial u_z\over\partial z},
\end{equation}
\begin{equation}\label{stssstrnCYL3}
\sigma_{zz}=\lambda{\partial u_r\over\partial r}+{\lambda\over r}\left({\partial u_\theta\over\partial \theta}+u_r\right)+(\lambda+2\mu){\partial u_z\over\partial z},
\end{equation}
\begin{equation}\label{stssstrnCYL4}
\sigma_{r \theta}=\mu\left({1\over r}{\partial u_r\over\partial \theta}+{\partial u_\theta\over\partial r}-{u_\theta\over r}\right)=\sigma_{\theta r},
\end{equation}
\begin{equation}\label{stssstrnCYL5}
\sigma_{rz}=\mu\left({\partial u_r\over\partial z}+{\partial u_z\over\partial r}\right)=\sigma_{zr},
\end{equation}
\begin{equation}\label{stssstrnCYL6}
\sigma_{\theta z}=\mu\left({\partial u_\theta\over\partial z}+{1\over r}{\partial u_z\over\partial \theta}\right)=\sigma_{z\theta}.
\end{equation}
\end{subequations}
Relations (\ref{strssdispCYL}), which provide the general mathematical connection between the components of the displacement and stress fields, will be used extensively in solving the above-defined boundary-value problem. 

Since the excitations are time-harmonic, relations (\ref{strssdispCYL}) and solution uniqueness together imply that the response of the cylinder must necessarily be so as well. In other words, the displacement field $\mathbf{u}(r,\theta,z,t)=\widetilde{\mathbf{u}}(r,\theta,z)\sin(\omega t)$, where $\widetilde{\mathbf{u}}(r,\theta,z)$ denotes the stationary or time-independent part of the displacement field. It is the latter object that we ultimately seek to determine and to then study.

\section{Some Parametric Solutions to the Navier-Lam\'{e} Equation}\label{SOLNSgen} 

In the absence of body forces, the following parametric solutions to Eq.~(\ref{NLE}) can be obtained using a Buchwald decomposition of the displacement field (see Ref.~\cite{usB1} for details): 
\begin{eqnarray}\label{solnradcomp}
u_r&=&\left(\sum_{s=1}^2\left[a_s\left \{ \begin{array}{c}
             J_n'(\alpha_sr) \\
             I_n'(\alpha_sr)
           \end{array} \right\} + b_s \left \{ \begin{array}{c}
             Y_n'(\alpha_sr) \\
             K_n'(\alpha_sr)
           \end{array} \right\}\right] \Big[c_s\cos(n\theta)+d_s\sin(n\theta)\Big]\right)\phi_z(z)\phi_t(t) \nonumber \\ \nonumber \\
           &+&{n\over r}\left[a_3\left \{ \begin{array}{c}
             J_n(\alpha_2r) \\
             I_n(\alpha_2r)
           \end{array} \right\} + b_3 \left \{ \begin{array}{c}
             Y_n(\alpha_2r) \\
             K_n(\alpha_2r)
           \end{array} \right\}\right] \Big[-c_3\sin(n\theta)+d_3\cos(n\theta)\Big]\chi^{}_z(z)\chi^{}_t(t),   
\end{eqnarray}

\begin{eqnarray}\label{solnangcomp}
u_\theta&=&{n\over r}\left(\sum_{s=1}^2\left[a_s\left \{ \begin{array}{c}
             J_n(\alpha_sr) \\
             I_n(\alpha_sr)
           \end{array} \right\} + b_s \left \{ \begin{array}{c}
             Y_n(\alpha_sr) \\
             K_n(\alpha_sr)
           \end{array} \right\}\right] \Big[-c_s\sin(n\theta)+d_s\cos(n\theta)\Big]\right)\phi_z(z)\phi_t(t) \nonumber \\ \nonumber \\
           &-&\left[a_3\left \{ \begin{array}{c}
             J_n'(\alpha_2r) \\
             I_n'(\alpha_2r)
           \end{array} \right\} + b_3 \left \{ \begin{array}{c}
             Y_n'(\alpha_2r) \\
             K_n'(\alpha_2r)
           \end{array} \right\}\right] \Big[c_3\cos(n\theta)+d_3\sin(n\theta)\Big]\chi^{}_z(z)\chi^{}_t(t),   
\end{eqnarray}
and
\begin{eqnarray}\label{solnaxcomp}
u_z=\left(\sum_{s=1}^2~\gamma_s\left[a_s\left \{ \begin{array}{c}
             J_n(\alpha_sr) \\
             I_n(\alpha_sr)
           \end{array} \right\} + b_s \left \{ \begin{array}{c}
             Y_n(\alpha_sr) \\
             K_n(\alpha_sr)
           \end{array} \right\}\right] \Big[c_s\cos(n\theta)+d_s\sin(n\theta)\Big]\right){\text{d}\psi_z(z)\over\text{d}z}~\psi_t(t), \nonumber \\
\end{eqnarray}
where $n$ is a non-negative integer and $\{a_1,a_2,a_3,b_1,b_2,b_3,c_1,c_2,c_3,d_1,d_2,d_3\}$ are arbitrary constants. The constituents of Eqs.~(\ref{solnradcomp})-(\ref{solnaxcomp}) are as follows:

\noindent \textbf{(i)} The constants $\alpha_1$ and $\alpha_2$ in the arguments of the Bessel functions are given by 
\begin{equation}\label{alphas}
\alpha_1=\sqrt{~\left|\kappa-{\rho\tau\over(\lambda+2\mu)}\right|~}, \quad \alpha_2=\sqrt{~\left|\kappa-{\rho\tau\over\mu}\right|~},
\end{equation}
where $\displaystyle \kappa\in\mathbb{R}\backslash\{0\}$ and $\displaystyle \tau\in\mathbb{R}\backslash\{0\}$ are free parameters. 

\noindent \textbf{(ii)} The correct linear combination of Bessel functions is determined by the relative values of the parameters $\{\lambda,\mu,\rho,\kappa,\tau\}$ as given in Table \ref{TabLinCombos}. 

\vspace*{0.25cm}

\setlength{\extrarowheight}{5pt}
\begin{table}[h]
\centering
\begin{tabular}{| c | c | c |}
    \hline
    ~~~Linear Combination~~~ & ~~~~~ $s=1$ term ~~~~~ & ~~~~~$s=2$ term ~~~~~\\ [10pt] \hline     
    $\displaystyle\{J_n(\alpha_sr), Y_n(\alpha_sr)\}$ & $\displaystyle \kappa > {\rho\tau\over(\lambda+2\mu)}$ & $\displaystyle \kappa > {\rho\tau\over\mu}$ \\ [10pt] \hline
    $\displaystyle\{I_n(\alpha_sr), K_n(\alpha_sr)\}$ & $\displaystyle \kappa < {\rho\tau\over(\lambda+2\mu)}$ & $\displaystyle \kappa < {\rho\tau\over\mu}$ \\ [10pt] \hline
\end{tabular}
\caption{Conditions on the radial part of each term in Eqs.~(\ref{solnradcomp})-(\ref{solnaxcomp}).}
\label{TabLinCombos}
\end{table}
\setlength{\extrarowheight}{1pt}

\noindent \textbf{(iii)} In Eqs.~(\ref{solnradcomp})-(\ref{solnangcomp}), primes denote differentiation with respect to the radial coordinate $r$.

\noindent \textbf{(iv)} The functions $\phi_z(z)$, $\phi_t(t)$, $\psi_z(z)$, $\psi_t(t)$, $\chi^{}_z(z)$, and $\chi^{}_t(t)$ are given by 
\begin{eqnarray}\label{Zpart}
\phi_z(z)=\psi_z(z)=\left \{ \begin{array}{lr}
             E\cos\left(\sqrt{|\kappa|}z\right)+F\sin\left(\sqrt{|\kappa|}z\right) & \text{if}~\kappa<0 \\
             E\exp\left(-\sqrt{\kappa}z\right)+F\exp\left(\sqrt{\kappa}z\right) & \text{if}~\kappa>0
           \end{array} \right.,
\end{eqnarray}

\begin{eqnarray}\label{Tpart}
\phi_t(t)=\psi_t(t)=\left \{ \begin{array}{lr}
             G\cos\left(\sqrt{|\tau|}t\right)+H\sin\left(\sqrt{|\tau|}t\right) & \text{if}~\tau<0 \\
             G\exp\left(-\sqrt{\tau}t\right)+H\exp\left(\sqrt{\tau}t\right) & \text{if}~\tau>0
           \end{array} \right.,
\end{eqnarray}

\begin{eqnarray}\label{ZpartC}
\chi^{}_z(z)=\left \{ \begin{array}{lr}
             \widetilde{E}\cos\left(\sqrt{|\kappa|}z\right)+\widetilde{F}\sin\left(\sqrt{|\kappa|}z\right) & \text{if}~\kappa<0 \\
             \widetilde{E}\exp\left(-\sqrt{\kappa}z\right)+\widetilde{F}\exp\left(\sqrt{\kappa}z\right) & \text{if}~\kappa>0
           \end{array} \right.,
\end{eqnarray}

\begin{eqnarray}\label{TpartC}
\chi^{}_t(t)=\left \{ \begin{array}{lr}
             \widetilde{G}\cos\left(\sqrt{|\tau|}t\right)+\widetilde{H}\sin\left(\sqrt{|\tau|}t\right) & \text{if}~\tau<0 \\
             \widetilde{G}\exp\left(-\sqrt{\tau}t\right)+\widetilde{H}\exp\left(\sqrt{\tau}t\right) & \text{if}~\tau>0
           \end{array} \right.,
\end{eqnarray}
where $\displaystyle\left\{E,F,G,H,\widetilde{E},\widetilde{F},\widetilde{G},\widetilde{H}\right\}$ are arbitrary constants. 

\noindent \textbf{(v)} The constant $\gamma_s$ in Eq.~(\ref{solnaxcomp}) is given by 
\begin{eqnarray}\label{solpsiPP2}
\gamma_s=\left \{ \begin{array}{lr}
             1 & \text{if}~s=1 \\
             {1\over\kappa}\left(\kappa - {\rho\tau\over\mu}\right) & \text{if}~s=2
           \end{array} \right.. 
\end{eqnarray}

Note that Eqs.~(\ref{solnradcomp})-(\ref{solnaxcomp}) are valid so long as $(\lambda+2\mu)\kappa \neq {\rho\tau}$ (i.e., $\alpha_1\neq0$) and $\mu\kappa \neq {\rho\tau}$ (i.e., $\alpha_2\neq0$); otherwise the radial parts must be modified as discussed in Ref.~\cite{usB1}. In the following, these conditions will be satisfied, by construction. 

\section{General Form of the Displacement Field}\label{GenFormDispField}

\subsection{The General Case $\displaystyle k\neq0$}\label{generalKneq0}

When $k\neq0$, general solutions suited to the boundary-value problem defined in Section~\ref{EGFULL} can be easily constructed from the parametric solutions given in Section~\ref{SOLNSgen} by identifying one or a combination of the physical parameters $\{L,R_1,R_2,k,\omega\}$ with the (free) mathematical parameters $\kappa$ and $\tau$. 
Let $\kappa=-\left({k\pi\over L}\right)^2$ and $\tau=-\omega^2$, and then choose particular solutions defined by taking $E=0$ in Eq.~(\ref{Zpart}), $\widetilde{E}=0$ in Eq.~(\ref{ZpartC}), $G=0$ in Eq.~(\ref{Tpart}), $\widetilde{G}=0$ in Eq.~(\ref{TpartC}), and $n=m$ in Eqs.~(\ref{solnradcomp})-(\ref{solnaxcomp}). Noting the forms of Eqs.~(\ref{solnradcomp})-(\ref{solnaxcomp}) and comparing (\ref{strssrr2}) with (\ref{stssstrnCYL1}), (\ref{strssrt2}) with (\ref{stssstrnCYL4}), and (\ref{strssrz2}) with (\ref{stssstrnCYL5}), we may immediately deduce that the axial and temporal parts of the displacement components are given by 
\begin{equation}\label{phipsiZEG}
\phi_z(z)=\psi_z(z)=F\sin\left({k\pi\over L}z\right), \quad \chi^{}_z(z)=\widetilde{F}\sin\left({k\pi\over L}z\right), 
\end{equation}
\begin{equation}\label{phipsiTEG}
\phi_t(t)=\psi_t(t)=H\sin(\omega t), \quad \chi^{}_t(t)=\widetilde{H}\sin(\omega t).
\end{equation}
By defining a new set of arbitrary constants
\begin{subequations}\label{arbconsts}
\begin{equation}\label{consts1}
\bar{A}_s\equiv a_sc_sFH,~\bar{B}_s\equiv b_sc_sFH,~\widetilde{A}_s\equiv a_sd_sFH,~\widetilde{B}_s\equiv b_sd_sFH, \quad (s=1,2)
\end{equation}
\begin{equation}\label{consts3}
\bar{A}_3\equiv a_3d_3\widetilde{F}\widetilde{H},~\bar{B}_3\equiv b_3d_3\widetilde{F}\widetilde{H},~\widetilde{A}_3\equiv a_3c_3\widetilde{F}\widetilde{H},~\widetilde{B}_3\equiv b_3c_3\widetilde{F}\widetilde{H},
\end{equation}
\end{subequations}
the following two independent particular solutions may be extracted from Eqs.~(\ref{solnradcomp})-(\ref{solnaxcomp}):
\begin{subequations}\label{GenSolA}
\begin{equation}\label{SolA1}
u_r=\Bigg(\sum_{s=1}^2\bigg[\widetilde{A}_s\Big\{\text{\large\textcircled{\normalsize 1}}\Big\}+\widetilde{B}_s\Big\{\text{\large\textcircled{\normalsize 2}}\Big\}\bigg]-{m\over r}\bigg[\widetilde{A}_3\Big\{\text{\large\textcircled{\normalsize 3}}\Big\}+\widetilde{B}_3\Big\{\text{\large\textcircled{\normalsize 4}}\Big\}\bigg]\Bigg)\sin(m\theta)\sin\left({k\pi\over L}z\right)\sin(\omega t),   
\end{equation}
\begin{equation}\label{SolA2}
u_\theta=\Bigg({m\over r}\left(\sum_{s=1}^2\bigg[\widetilde{A}_s\Big\{\text{\large\textcircled{\normalsize 5}}\Big\}+\widetilde{B}_s\Big\{\text{\large\textcircled{\normalsize 6}}\Big\}\bigg]\right)-\bigg[\widetilde{A}_3\Big\{\text{\large\textcircled{\normalsize 7}}\Big\}+\widetilde{B}_3\Big\{\text{\large\textcircled{\normalsize 8}}\Big\}\bigg]\Bigg)\cos(m\theta)\sin\left({k\pi\over L}z\right)\sin(\omega t),   
\end{equation}
\begin{equation}\label{SolA3}
u_z=\left({k\pi\over L}\right)\Bigg(\sum_{s=1}^2\gamma_s\bigg[\widetilde{A}_s\Big\{\text{\large\textcircled{\normalsize 9}}\Big\}+\widetilde{B}_s\Big\{\text{\Large\textcircled{\normalsize 10}}\Big\}\bigg]\Bigg)\sin(m\theta)\cos\left({k\pi\over L}z\right)\sin(\omega t),   
\end{equation}
\end{subequations}
and
\begin{subequations}\label{GenSolB}
\begin{equation}\label{SolB1}
u_r=\Bigg(\sum_{s=1}^2\bigg[\bar{A}_s\Big\{\text{\large\textcircled{\normalsize 1}}\Big\}+\bar{B}_s\Big\{\text{\large\textcircled{\normalsize 2}}\Big\}\bigg]+{m\over r}\bigg[\bar{A}_3\Big\{\text{\large\textcircled{\normalsize 3}}\Big\}+\bar{B}_3\Big\{\text{\large\textcircled{\normalsize 4}}\Big\}\bigg]\Bigg)\cos(m\theta)\sin\left({k\pi\over L}z\right)\sin(\omega t),   
\end{equation}
\begin{equation}\label{SolB2}
u_\theta=-\Bigg({m\over r}\left(\sum_{s=1}^2\bigg[\bar{A}_s\Big\{\text{\large\textcircled{\normalsize 5}}\Big\}+\bar{B}_s\Big\{\text{\large\textcircled{\normalsize 6}}\Big\}\bigg]\right)+\bar{A}_3\Big\{\text{\large\textcircled{\normalsize 7}}\Big\}+\bar{B}_3\Big\{\text{\large\textcircled{\normalsize 8}}\Big\}\Bigg)\sin(m\theta)\sin\left({k\pi\over L}z\right)\sin(\omega t),   
\end{equation}
\begin{equation}\label{SolB3}
u_z=\left({k\pi\over L}\right)\Bigg(\sum_{s=1}^2\gamma_s\bigg[\bar{A}_s\Big\{\text{\large\textcircled{\normalsize 9}}\Big\}+\bar{B}_s\Big\{\text{\Large\textcircled{\normalsize 10}}\Big\}\bigg]\Bigg)\cos(m\theta)\cos\left({k\pi\over L}z\right)\sin(\omega t),   
\end{equation}
\end{subequations}
where
\begin{subequations}\label{radpartsGEN}
\begin{eqnarray}\label{radparts1}
\Big\{\text{\large\textcircled{\normalsize 1}}\Big\}=\left \{ \begin{array}{c}
             J_m'(\alpha_sr)={m\over r}J_m(\alpha_sr)-\alpha_sJ_{m+1}(\alpha_sr) \\
             I_m'(\alpha_sr)={m\over r}I_m(\alpha_sr)+\alpha_sI_{m+1}(\alpha_sr)
           \end{array} \right\},
\end{eqnarray}
\begin{eqnarray}\label{radparts2}
\Big\{\text{\large\textcircled{\normalsize 2}}\Big\}=\left \{ \begin{array}{c}
             Y_m'(\alpha_sr)={m\over r}Y_m(\alpha_sr)-\alpha_sY_{m+1}(\alpha_sr) \\
             K_m'(\alpha_sr)={m\over r}K_m(\alpha_sr)-\alpha_sK_{m+1}(\alpha_sr)
           \end{array} \right\},
\end{eqnarray}
\begin{eqnarray}\label{radparts3}
\Big\{\text{\large\textcircled{\normalsize 3}}\Big\}=\left \{ \begin{array}{c}
             J_m(\alpha_2r) \\
             I_m(\alpha_2r) 
           \end{array} \right\}, \quad \Big\{\text{\large\textcircled{\normalsize 4}}\Big\}=\left \{ \begin{array}{c}
             Y_m(\alpha_2r) \\
             K_m(\alpha_2r) 
           \end{array} \right\},
\end{eqnarray}
\begin{eqnarray}\label{radparts4}
\Big\{\text{\large\textcircled{\normalsize 5}}\Big\}=\Big\{\text{\large\textcircled{\normalsize 9}}\Big\}=\left \{ \begin{array}{c}
             J_m(\alpha_sr) \\
             I_m(\alpha_sr) 
           \end{array} \right\}, \quad \Big\{\text{\large\textcircled{\normalsize 6}}\Big\}=\Big\{\text{\Large\textcircled{\normalsize 10}}\Big\}=\left \{ \begin{array}{c}
             Y_m(\alpha_sr) \\
             K_m(\alpha_sr) 
           \end{array} \right\},
\end{eqnarray}
\begin{eqnarray}\label{radparts5}
\Big\{\text{\large\textcircled{\normalsize 7}}\Big\}=\left \{ \begin{array}{c}
             J_m'(\alpha_2r)={m\over r}J_m(\alpha_2r)-\alpha_2J_{m+1}(\alpha_2r) \\
             I_m'(\alpha_2r)={m\over r}I_m(\alpha_2r)+\alpha_2I_{m+1}(\alpha_2r)
           \end{array} \right\},
\end{eqnarray}
\begin{eqnarray}\label{radparts6}
\Big\{\text{\large\textcircled{\normalsize 8}}\Big\}=\left \{ \begin{array}{c}
             Y_m'(\alpha_2r)={m\over r}Y_m(\alpha_2r)-\alpha_2Y_{m+1}(\alpha_2r) \\
             K_m'(\alpha_2r)={m\over r}K_m(\alpha_2r)-\alpha_2K_{m+1}(\alpha_2r)
           \end{array} \right\}.
\end{eqnarray}
\end{subequations}

Note that particular solutions (\ref{GenSolA}) and (\ref{GenSolB}) automatically satisfy end conditions (\ref{SSBCs1}) and (\ref{SSBCs2}). Note also that (by virtue of (\ref{stssstrnCYL3})) $\sigma_{zz}(r,\theta,z,t)=F(r,\theta)\sin\left({k\pi\over L}z\right)\sin(\omega t)$, where the precise form of $F(r,\theta)$ is not relevant for our purposes, and thus displacements (\ref{GenSolA}) and (\ref{GenSolB}) as well automatically satisfy end conditions (\ref{SSBCs3}). It can be deduced from inspection of (\ref{strssdispCYL}), (\ref{GenSolA}), and (\ref{GenSolB}), that solution (\ref{GenSolA}) is incompatible with boundary conditions (\ref{strssrr2})-(\ref{strssrz2}), whereas solution (\ref{GenSolB}) is compatible. Solution (\ref{GenSolB}) therefore furnishes the general form of the displacement field appropriate to the boundary-value problem defined in Section~\ref{EGFULL}. Henceforth, we switch to a less cumbersome notation by dropping the bars above the arbitrary constants.  

The proper choices of Bessel functions in the radial parts of the displacement components depend on the relative values of the material and excitation parameters; three cases can be distinguished as listed in Table \ref{EG1Cases}. In order to provide a complete solution that involves only real-valued Bessel functions, the problem will be solved separately for each of these three cases. Note that there are two degenerate cases not included in Table \ref{EG1Cases}: (i) ${\rho\omega^2\over(\lambda+2\mu)}=\left({k\pi\over L}\right)^2$; and (ii) ${\rho\omega^2\over\mu}=\left({k\pi\over L}\right)^2$. These two cases require special treatment and shall not be considered here. As a final remark, note that solution (\ref{GenSolB}) is not valid when $k=0$. 

\setlength{\extrarowheight}{10pt}
\begin{table}[h]
\centering
\begin{tabular}{| c | c | c |} \hline 
    ~Case~ & ~Parametric Relationship ($k$, $\omega$)~ & ~Parametric Relationship ($\kappa$, $\tau$)~ \\ [5pt] \hline 
     1 & $\displaystyle {\rho\omega^2\over(\lambda+2\mu)}<{\rho\omega^2\over\mu}<\left({k\pi\over L}\right)^2$ & $\displaystyle \kappa<{\rho\tau\over(\lambda+2\mu)}$ and $\displaystyle \kappa<{\rho\tau\over\mu}$ \\ [10pt] \hline 
     2 & $\displaystyle \left({k\pi\over L}\right)^2<{\rho\omega^2\over(\lambda+2\mu)}<{\rho\omega^2\over\mu}$ & $\displaystyle \kappa>{\rho\tau\over(\lambda+2\mu)}$ and $\displaystyle \kappa>{\rho\tau\over\mu}$ \\ [10pt] \hline 
     3 & $\displaystyle {\rho\omega^2\over(\lambda+2\mu)}<\left({k\pi\over L}\right)^2<{\rho\omega^2\over\mu}$ & $\displaystyle \kappa<{\rho\tau\over(\lambda+2\mu)}$ and $\displaystyle \kappa>{\rho\tau\over\mu}$ \\ [10pt] \hline 
\end{tabular}
\caption{Parametric relationships defining three distinct sub-problems. In the second column, the relationship is expressed in terms of the physical excitation parameters $k$ and $\omega$, whereas in the third column, the relationship is expressed in terms of the mathematical parameters $\kappa$ and $\tau$.}
\label{EG1Cases}
\end{table} 
\setlength{\extrarowheight}{1pt}

\subsection{The Special Case $\displaystyle k=0$}\label{generalKeq0}

It can be established (employing results from Ref.~\cite{usB1} or otherwise) that 
\begin{equation}\label{gensolnkeq0}
u_r=0, \quad u_\theta=0, \quad u_z=\Big[AJ_m(\alpha r)+BY_m(\alpha r)\Big]\cos(m\theta)\sin(\omega t), 
\end{equation}
where $\alpha=\sqrt{\rho\omega^2/\mu}$, is a solution to Eq.~(\ref{NLE}) that is furthermore compatible with boundary conditions (\ref{strssrr2})-(\ref{strssrz2}) when $k=0$. Solution (\ref{gensolnkeq0}) therefore furnishes the general form of the displacement field appropriate to the boundary-value problem defined in Section~\ref{EGFULL} in the special case $k=0$.

\section{Analytics I: General Case $\displaystyle k\neq0$}\label{solntoBVP}

\subsection{\textbf{Case 1:} $\displaystyle{\rho\omega^2\over(\lambda+2\mu)}<{\rho\omega^2\over\mu}<\left({k\pi\over L}\right)^2$}\label{BVP2C1}

In this case, $\kappa<{\rho\tau/(\lambda+2\mu)}$ and $\kappa<{\rho\tau/\mu}$ (c.f., Table \ref{EG1Cases}), and thus, according to Table \ref{TabLinCombos}, linear combinations of $\displaystyle\{I_m(\alpha_sr), K_m(\alpha_sr)\}$ (and their derivatives) should be employed in the radial parts of (\ref{GenSolB}) (i.e., the modified Bessel functions should be chosen from  (\ref{radpartsGEN})), where the constants $\alpha_1$ and $\alpha_2$, as determined from Eq.~(\ref{alphas}), are:
\begin{equation}\label{alphasEG}
\alpha_1=\sqrt{\left({k\pi\over L}\right)^2-{\rho\omega^2\over(\lambda+2\mu)}}, \quad \alpha_2=\sqrt{\left({k\pi\over L}\right)^2-{\rho\omega^2\over\mu}}.
\end{equation}
The constant $\gamma_s$ in Eq.~(\ref{SolB3}), as determined from Eq.~(\ref{solpsiPP2}), is given by 
\begin{eqnarray}\label{solpsiPP2EG}
\displaystyle
\gamma_s=\left \{ \begin{array}{lr}
             1 & ~~\text{if}~s=1 \\
             1-\left[\left({\rho\omega^2\over\mu}\right)/\left({k\pi\over L}\right)^2\right] & ~~\text{if}~s=2
           \end{array} \right..
\end{eqnarray}
Inputting the above ingredients into (\ref{GenSolB}), the displacement components take the form: 
\begin{subequations}\label{SolBC1}
\begin{eqnarray}\label{SolBC11}
u_r&=&\Bigg\{\sum_{s=1}^2\bigg[A_s\bigg({m\over r}I_m(\alpha_sr)+\alpha_sI_{m+1}(\alpha_sr)\bigg)+B_s\bigg({m\over r}K_m(\alpha_sr)-\alpha_sK_{m+1}(\alpha_sr)\bigg)\bigg] \nonumber \\ 
&&~~+~{m\over r}\Big[A_3I_m(\alpha_2r)+B_3K_m(\alpha_2r)\Big]\Bigg\}\cos(m\theta)\sin\left({k\pi\over L}z\right)\sin(\omega t),   
\end{eqnarray}
\begin{eqnarray}\label{SolBC12}
u_\theta&=&-\Bigg\{{m\over r}\left(\sum_{s=1}^2\Big[A_sI_m(\alpha_sr)+B_sK_m(\alpha_sr)\Big]\right)+A_3\bigg[{m\over r}I_m(\alpha_2r)+\alpha_2I_{m+1}(\alpha_2r)\bigg] \nonumber \\ 
&&~~~~+~B_3\bigg[{m\over r}K_m(\alpha_2r)-\alpha_2K_{m+1}(\alpha_2r)\bigg]\Bigg\}\sin(m\theta)\sin\left({k\pi\over L}z\right)\sin(\omega t),   
\end{eqnarray}
\begin{eqnarray}\label{SolBC13}
u_z=\left({k\pi\over L}\right)\Bigg\{\sum_{s=1}^2\gamma_s\Big[A_sI_m(\alpha_sr)+B_sK_m(\alpha_sr)\Big]\Bigg\}\cos(m\theta)\cos\left({k\pi\over L}z\right)\sin(\omega t),   
\end{eqnarray}
\end{subequations}
where the constants $\alpha_s$ and $\gamma_s$ are given by Eqs.~(\ref{alphasEG}) and (\ref{solpsiPP2EG}), respectively. 

We must now determine the values of the constants $\displaystyle\left\{A_1,A_2,A_3,B_1,B_2,B_3\right\}$ in Eqs.~(\ref{SolBC11})-(\ref{SolBC13}) that satisfy boundary conditions (\ref{strssrr2})-(\ref{strssrz2}). Substituting Eqs.~(\ref{SolBC11})-(\ref{SolBC13}) into Eqs.~(\ref{stssstrnCYL1}), (\ref{stssstrnCYL4}), and (\ref{stssstrnCYL5}), and performing the lengthy algebra yields the stress components:
\begin{subequations}\label{strsscmpRRC1}
\begin{eqnarray}\label{strsscmpRRC11A}
\sigma_{rr}(r,\theta,z,t)&=&2\mu\Bigg\{\sum_{s=1}^2A_s\left[\left({\beta_s\over2\mu}+{m(m-1)\over r^2}\right)I_m(\alpha_sr)-{\alpha_s\over r}I_{m+1}(\alpha_sr)\right] \nonumber \\
&&~~+~A_3\left[{m(m-1)\over r^2}I_m(\alpha_2r)+{\alpha_2m\over r}I_{m+1}(\alpha_2r)\right] \nonumber \\ 
&&~~+~\sum_{s=1}^2B_s\left[\left({\beta_s\over2\mu}+{m(m-1)\over r^2}\right)K_m(\alpha_sr)+{\alpha_s\over r}K_{m+1}(\alpha_sr)\right] \nonumber \\ 
&&~~+~B_3\left[{m(m-1)\over r^2}K_m(\alpha_2r)-{\alpha_2m\over r}K_{m+1}(\alpha_2r)\right]\Bigg\} \nonumber \\ 
&&\quad \quad \ \times \cos(m\theta)\sin\left({k\pi\over L}z\right)\sin(\omega t),
\end{eqnarray}
where
\begin{equation}\label{strsscmpRRC11B}
\beta_s=\lambda\left[\alpha^2_s-\gamma_s\left({k\pi\over L}\right)^2\right]+2\mu\alpha^2_s, \quad s=1,2 
\end{equation}
\end{subequations} 
\begin{eqnarray}\label{strsscmpRTC1}
\sigma_{r\theta}(r,\theta,z,t)&=&-2\mu\Bigg\{\sum_{s=1}^2A_s\bigg[{m(m-1)\over r^2}I_m(\alpha_sr)+{\alpha_sm\over r}I_{m+1}(\alpha_sr)\bigg] \nonumber \\ 
&&~~~~~+~A_3\left[\bigg({\alpha_2^2\over2}+{m(m-1)\over r^2}\bigg)I_m(\alpha_2r)-{\alpha_2\over r}I_{m+1}(\alpha_2r)\right] \nonumber \\ 
&&~~~~~+~\sum_{s=1}^2B_s\bigg[{m(m-1)\over r^2}K_m(\alpha_sr)-{\alpha_sm\over r}K_{m+1}(\alpha_sr)\bigg] \nonumber \\ 
&&~~~~~+~B_3\left[\bigg({\alpha_2^2\over2}+{m(m-1)\over r^2}\bigg)K_m(\alpha_2r)+{\alpha_2\over r}K_{m+1}(\alpha_2r)\right] \Bigg\} \nonumber \\
&& \quad \quad \quad \times \sin(m\theta)\sin\left({k\pi\over L}z\right)\sin(\omega t),
\end{eqnarray}
and 
\begin{eqnarray}\label{strsscmpRZC1}
\sigma_{rz}(r,\theta,z,t)&=&\mu\left({k\pi\over L}\right)\Bigg\{\sum_{s=1}^2A_s\left(1+\gamma_s\right)\bigg[{m\over r}I_m(\alpha_sr)+\alpha_sI_{m+1}(\alpha_sr)\bigg] \nonumber \\
&&+~A_3\bigg[{m\over r}I_m(\alpha_2r)\bigg]+\sum_{s=1}^2B_s\left(1+\gamma_s\right)\bigg[{m\over r}K_m(\alpha_sr)-\alpha_sK_{m+1}(\alpha_sr)\bigg] \nonumber \\ 
&&+~B_3\bigg[{m\over r}K_m(\alpha_2r)\bigg]\Bigg\}\cos(m\theta)\cos\left({k\pi\over L}z\right)\sin(\omega t).
\end{eqnarray}

Application of the boundary conditions then proceeds by substituting Eqs.~(\ref{strsscmpRRC1}), (\ref{strsscmpRTC1}), and (\ref{strsscmpRZC1}) into the LHSs of Eqs.~(\ref{strssrr2}), (\ref{strssrt2}), and (\ref{strssrz2}), respectively, and then canceling identical sinusoidal terms on both sides of the resulting equations. When $m\neq0$, this procedure yields six conditions that can be compactly written as the following $6\times6$ linear system:
\begin{subequations}\label{BCmatEQC1}
\begin{eqnarray}\label{BCmatEQC11}
{\resizebox{0.45\textwidth}{!}{$\left[\begin{array}{c|c}
~\mathbf{A}_1~ & ~\mathbf{B}_1~ \\ 
\hline
~\mathbf{A}_2~ & ~\mathbf{B}_2~
\end{array}\right] \left[\begin{array}{c}
            \mathbf{X}_A \\
            \hline
            \mathbf{X}_B 
           \end{array} \right] =  \left[\begin{array}{r}
           \mathbf{S}_1 \\
           \hline
           \mathbf{S}_2     
           \end{array} \right]$}},
\end{eqnarray}
where, using a shorthand notation, the $3\times3$ matrix blocks $\displaystyle\{\mathbf{A}_i,\mathbf{B}_i:i=1,2\}$ are  
\begin{eqnarray}\label{BCmatEQC11A}
\mathbf{A}_i=\left[\begin{array}{ccc}
            f_{m,i}-v_{m{+}1,i} & g_{m,i}-w_{m{+}1,i} & (m{-}1)q_{m,i}+mw_{m{+}1,i} \\
            (m{-}1)p_{m,i}+mv_{m{+}1,i} & ~(m{-}1)q_{m,i}+mw_{m{+}1,i}~ & h_{m,i}-w_{m{+}1,i} \\
            2(p_{m,i}+v_{m{+}1,i}) & (1{+}\gamma_2)(q_{m,i}+w_{m{+}1,i}) & q_{m,i}  
            \end{array} \right], 
\end{eqnarray}
\begin{eqnarray}\label{BCmatEQC11B}
\mathbf{B}_i=\left[\begin{array}{ccc}
\widetilde{f}_{m,i}+\widetilde{v}_{m{+}1,i} & \widetilde{g}_{m,i}+\widetilde{w}_{m{+}1,i} & (m{-}1)\widetilde{q}_{m,i}-m\widetilde{w}_{m{+}1,i} \\
(m{-}1)\widetilde{p}_{m,i}-m\widetilde{v}_{m{+}1,i} & ~(m{-}1)\widetilde{q}_{m,i}-m\widetilde{w}_{m{+}1,i}~ & \widetilde{h}_{m,i}+\widetilde{w}_{m{+}1,i} \\
2(\widetilde{p}_{m,i}-\widetilde{v}_{m{+}1,i}) & (1{+}\gamma_2)(\widetilde{q}_{m,i}-\widetilde{w}_{m{+}1,i}) & \widetilde{q}_{m,i} \\
\end{array} \right], 
\end{eqnarray}
and the $3\times1$ column blocks $\displaystyle\{\mathbf{X}_A,\mathbf{X}_B, \mathbf{S}_1, \mathbf{S}_2\}$ are 
\begin{eqnarray}\label{BCmatEQC11C}
\mathbf{X}_A=\left[\begin{array}{c}
            A_1 \\
            A_2 \\
            A_3 
           \end{array} \right], \quad \mathbf{X}_B=\left[\begin{array}{c}
            B_1 \\
            B_2 \\
            B_3 
           \end{array} \right], \quad \mathbf{S}_1=\left[\begin{array}{r}
           \mathbb{A}~ \\
           -\mathbb{B}~ \\
           \mathbb{C}~ 
           \end{array} \right], \quad \mathbf{S}_2=\left[\begin{array}{r}
           \mathbb{D}~ \\
           -\mathbb{E}~ \\
           \mathbb{F}~     
           \end{array} \right].
\end{eqnarray}
The shorthand notation employed in Eqs.~(\ref{BCmatEQC11A})-(\ref{BCmatEQC11C}) is as follows:
\begin{eqnarray}\label{BCmatEQC12a}
\left\{\begin{array}{c} f_{m,i} \\ \widetilde{f}_{m,i} \end{array} \right\} \equiv \left({\beta_1R_i\over2\mu}+{m(m-1)\over R_i}\right) \left\{\begin{array}{c} I_m(\alpha_1R_i) \\ K_m(\alpha_1R_i) \end{array} \right\}, \quad i=1,2 
\end{eqnarray}
\begin{eqnarray}\label{BCmatEQC12b}
\left\{\begin{array}{c} g_{m,i} \\ \widetilde{g}_{m,i} \end{array} \right\} \equiv \left({\beta_2R_i\over2\mu}+{m(m-1)\over R_i}\right) \left\{\begin{array}{c} I_m(\alpha_2R_i) \\ K_m(\alpha_2R_i) \end{array} \right\}, \quad i=1,2 
\end{eqnarray}
\begin{eqnarray}\label{BCmatEQC12c}
\left\{\begin{array}{c} h_{m,i} \\ \widetilde{h}_{m,i} \end{array} \right\} \equiv \left({\alpha^2_2R_i\over2}+{m(m-1)\over R_i}\right) \left\{\begin{array}{c} I_m(\alpha_2R_i) \\ K_m(\alpha_2R_i) \end{array} \right\}, \quad i=1,2 
\end{eqnarray}
\begin{eqnarray}\label{BCmatEQC13}
\left\{\begin{array}{c} p_{m,i} \\ \widetilde{p}_{m,i} \end{array} \right\} \equiv {m\over R_i} \left\{\begin{array}{c} I_m(\alpha_1R_i) \\ K_m(\alpha_1R_i) \end{array} \right\}, \quad i=1,2 
\end{eqnarray}
\begin{eqnarray}\label{BCmatEQC14}
\left\{\begin{array}{c} q_{m,i} \\ \widetilde{q}_{m,i} \end{array} \right\} \equiv {m\over R_i} \left\{\begin{array}{c} I_m(\alpha_2R_i) \\ K_m(\alpha_2R_i) \end{array} \right\}, \quad i=1,2 
\end{eqnarray}
\begin{eqnarray}\label{BCmatEQC15}
\left\{\begin{array}{c} v_{m{+}1,i} \\ \widetilde{v}_{m{+}1,i} \end{array} \right\} \equiv \alpha_1 \left\{\begin{array}{c} I_{m+1}(\alpha_1R_i) \\ K_{m+1}(\alpha_1R_i) \end{array} \right\}, \quad i=1,2 
\end{eqnarray}
\begin{eqnarray}\label{BCmatEQC16}
\left\{\begin{array}{c} w_{m{+}1,i} \\ \widetilde{w}_{m{+}1,i} \end{array} \right\} \equiv \alpha_2 \left\{\begin{array}{c} I_{m+1}(\alpha_2R_i) \\ K_{m+1}(\alpha_2R_i) \end{array} \right\}, \quad i=1,2 
\end{eqnarray}
and 
\begin{eqnarray}\label{BCmatEQC17}
\left[\mathbb{A}~~\mathbb{B}~~\mathbb{C}~~\mathbb{D}~~\mathbb{E}~~\mathbb{F}\right] \equiv \left[{\mathcal{A}R_1\over2\mu}~~{\mathcal{B}R_1\over2\mu}~~{\mathcal{C}L\over k\pi\mu}~~{\mathcal{D}R_2\over2\mu}~~{\mathcal{E}R_2\over2\mu}~~{\mathcal{F}L\over k\pi\mu}\right]. 
\end{eqnarray}
\end{subequations} 

\subsubsection{Special Case: $\displaystyle m=0$}

When $m=0$, $u_\theta=0$, $\sigma_{r\theta}=0$, and boundary conditions (\ref{strssrt2}) are identically satisfied. Application of boundary conditions (\ref{strssrr2}) and (\ref{strssrz2}) yields the $4\times4$ linear system: 
\begin{eqnarray}\label{BCmatEQC1SCm0}
\left[\begin{array}{cccc}
            f_{0,1}-v_{1,1} & g_{0,1}-w_{1,1} & \widetilde{f}_{0,1}+\widetilde{v}_{1,1} & \widetilde{g}_{0,1}+\widetilde{w}_{1,1} \\
            2v_{1,1} & ~(1{+}\gamma_2)w_{1,1}~ & -2\widetilde{v}_{1,1} & -(1{+}\gamma_2)\widetilde{w}_{1,1} \\  
            f_{0,2}-v_{1,2} & g_{0,2}-w_{1,2} & ~\widetilde{f}_{0,2}+\widetilde{v}_{1,2}~ & \widetilde{g}_{0,2}+\widetilde{w}_{1,2} \\
            2v_{1,2} & (1{+}\gamma_2)w_{1,2} & -2\widetilde{v}_{1,2} & -(1{+}\gamma_2)\widetilde{w}_{1,2} 
           \end{array} \right] \left[\begin{array}{c}
             A_1 \\
             A_2 \\
             B_1 \\ 
             B_2 
           \end{array} \right] =  \left[\begin{array}{c}
              \mathbb{A} \\
              \mathbb{C} \\ 
              \mathbb{D} \\
              \mathbb{F} 
           \end{array} \right],
\end{eqnarray}
where the matrix elements are the evaluated zero- and first-order Bessel functions obtained from substituting $m=0$ in Eqs.~(\ref{BCmatEQC12a}), (\ref{BCmatEQC12b}), (\ref{BCmatEQC15}), and (\ref{BCmatEQC16}). 

In this special case, the non-zero components of the displacement field reduce to:
\begin{subequations}\label{SolBVP2C1meq0}
\begin{equation}\label{SolBVP2C1meq01}
u_r(r,z,t)=\Bigg\{\sum_{s=1}^2\alpha_s\Big[A_sI_{1}(\alpha_sr)-B_sK_{1}(\alpha_sr)\Big]\Bigg\}\sin\left({k\pi\over L}z\right)\sin(\omega t), 
\end{equation}
\begin{equation}\label{SolBVP2C1meq02}
u_z(r,z,t)=\left({k\pi\over L}\right)\Bigg\{\sum_{s=1}^2\gamma_s\Big[A_sI_0(\alpha_sr)+B_sK_0(\alpha_sr)\Big]\Bigg\}\cos\left({k\pi\over L}z\right)\sin(\omega t),   
\end{equation}
\end{subequations}
where constants $\alpha_s$ and $\gamma_s$ are given by Eqs.~(\ref{alphasEG}) and (\ref{solpsiPP2EG}), respectively, and the constants $\left\{A_1,A_2,B_1,B_2\right\}$ are those obtained from solving Eq.~(\ref{BCmatEQC1SCm0}).  

\subsection{\textbf{Case 2:} $\displaystyle \left({k\pi\over L}\right)^2<{\rho\omega^2\over(\lambda+2\mu)}<{\rho\omega^2\over\mu}$}\label{BVP2C2}

According to Tables \ref{TabLinCombos} and \ref{EG1Cases}, the Bessel functions $\{J_m(\alpha_sr), Y_m(\alpha_sr)\}$ (and their derivatives) should in this case be employed in the radial parts of (\ref{GenSolB}) (i.e., the unmodified Bessel functions should be chosen from (\ref{radpartsGEN})). The  displacement components thus take the form:
\begin{subequations}\label{SolBC2}
\begin{eqnarray}\label{SolBC21}
u_r&=&\Bigg\{\sum_{s=1}^2\bigg[A_s\bigg({m\over r}J_m(\alpha_sr)-\alpha_sJ_{m+1}(\alpha_sr)\bigg)+B_s\bigg({m\over r}Y_m(\alpha_sr)-\alpha_sY_{m+1}(\alpha_sr)\bigg)\bigg] \nonumber \\ 
&&~~+~{m\over r}\Big[A_3J_m(\alpha_2r)+B_3Y_m(\alpha_2r)\Big]\Bigg\}\cos(m\theta)\sin\left({k\pi\over L}z\right)\sin(\omega t),   
\end{eqnarray}
\begin{eqnarray}\label{SolBC22}
u_\theta&=&-\Bigg\{{m\over r}\left(\sum_{s=1}^2\Big[A_sJ_m(\alpha_sr)+B_sY_m(\alpha_sr)\Big]\right)+A_3\bigg[{m\over r}J_m(\alpha_2r)-\alpha_2J_{m+1}(\alpha_2r)\bigg] \nonumber \\ 
&&~~~~+~B_3\bigg[{m\over r}Y_m(\alpha_2r)-\alpha_2Y_{m+1}(\alpha_2r)\bigg]\Bigg\}\sin(m\theta)\sin\left({k\pi\over L}z\right)\sin(\omega t),   
\end{eqnarray}
\begin{eqnarray}\label{SolBC23}
u_z=\left({k\pi\over L}\right)\Bigg\{\sum_{s=1}^2\gamma_s\Big[A_sJ_m(\alpha_sr)+B_sY_m(\alpha_sr)\Big]\Bigg\}\cos(m\theta)\cos\left({k\pi\over L}z\right)\sin(\omega t),   
\end{eqnarray}
\end{subequations}
where 
\begin{equation}\label{alphasEG5}
\alpha_1=\sqrt{-\left({k\pi\over L}\right)^2+{\rho\omega^2\over(\lambda+2\mu)}}, \quad \alpha_2=\sqrt{-\left({k\pi\over L}\right)^2+{\rho\omega^2\over\mu}},
\end{equation}
and $\gamma_s$ is again given by Eq.~(\ref{solpsiPP2EG}). 

The constants $\left\{A_1,A_2,A_3,B_1,B_2,B_3\right\}$ in Eqs.~(\ref{SolBC21})-(\ref{SolBC23}) must as before be chosen so as to satisfy boundary conditions (\ref{BCsCRVD2}). Proceeding as in the previous case, we first obtain general formulas for the radial components of the stress field. Substituting Eqs.~(\ref{SolBC21})-(\ref{SolBC23}) into Eqs.~(\ref{stssstrnCYL1}), (\ref{stssstrnCYL4}), and (\ref{stssstrnCYL5}), and performing the lengthy algebra yields the required stress components:
\begin{subequations}\label{strsscmpRRC2P1}
\begin{eqnarray}\label{strsscmpRRC2P1A}
\sigma_{rr}(r,\theta,z,t)&=&2\mu\Bigg\{\sum_{s=1}^2A_s\left[\left(-{\eta_s\over2\mu}+{m(m-1)\over r^2}\right)J_m(\alpha_sr)+{\alpha_s\over r}J_{m+1}(\alpha_sr)\right] \nonumber \\ 
&&~~+~A_3\left[{m(m-1)\over r^2}J_m(\alpha_2r)-{\alpha_2m\over r}J_{m+1}(\alpha_2r)\right] \nonumber \\
&&~~+\sum_{s=1}^2B_s\left[\left(-{\eta_s\over2\mu}+{m(m-1)\over r^2}\right)Y_m(\alpha_sr)+{\alpha_s\over r}Y_{m+1}(\alpha_sr)\right] \nonumber \\
&&~~+~B_3\left[{m(m-1)\over r^2}Y_m(\alpha_2r)-{\alpha_2m\over r}Y_{m+1}(\alpha_2r)\right] \Bigg\} \nonumber \\ 
&& \quad \quad \times \cos(m\theta)\sin\left({k\pi\over L}z\right)\sin(\omega t),
\end{eqnarray}
where
\begin{equation}\label{strsscmpRRC2P1B}
\eta_s=\lambda\left[\alpha^2_s+\gamma_s\left({k\pi\over L}\right)^2\right]+2\mu\alpha^2_s, \quad s=1,2 
\end{equation}
\end{subequations} 
\begin{eqnarray}\label{strsscmpRTC2P1}
\sigma_{r\theta}(r,\theta,z,t)&=&-2\mu\Bigg\{\sum_{s=1}^2A_s\bigg[{m(m-1)\over r^2}J_m(\alpha_sr)-{\alpha_sm\over r}J_{m+1}(\alpha_sr)\bigg] \nonumber \\ 
&&~~~~~+~A_3\left[\bigg({m(m-1)\over r^2}-{\alpha_2^2\over2}\bigg)J_m(\alpha_2r)+{\alpha_2\over r}J_{m+1}(\alpha_2r)\right] \nonumber \\ 
&&~~~~~+~\sum_{s=1}^2B_s\bigg[{m(m-1)\over r^2}Y_m(\alpha_sr)-{\alpha_sm\over r}Y_{m+1}(\alpha_sr)\bigg] \nonumber \\ 
&&~~~~~+~B_3\left[\bigg({m(m-1)\over r^2}-{\alpha_2^2\over2}\bigg)Y_m(\alpha_2r)+{\alpha_2\over r}Y_{m+1}(\alpha_2r)\right] \Bigg\} \nonumber \\
&& \quad \quad \quad \times \sin(m\theta)\sin\left({k\pi\over L}z\right)\sin(\omega t),
\end{eqnarray}
and 
\begin{eqnarray}\label{strsscmpRZC2P1}
\sigma_{rz}(r,\theta,z,t)&=&\mu\left({k\pi\over L}\right)\Bigg\{\sum_{s=1}^2A_s\left(1+\gamma_s\right)\bigg[{m\over r}J_m(\alpha_sr)-\alpha_sJ_{m+1}(\alpha_sr)\bigg] \nonumber \\
&&+~A_3\bigg[{m\over r}J_m(\alpha_2r)\bigg]+\sum_{s=1}^2B_s\left(1+\gamma_s\right)\bigg[{m\over r}Y_m(\alpha_sr)-\alpha_sY_{m+1}(\alpha_sr)\bigg] \nonumber \\ 
&&+~B_3\bigg[{m\over r}Y_m(\alpha_2r)\bigg]\Bigg\}\cos(m\theta)\cos\left({k\pi\over L}z\right)\sin(\omega t).
\end{eqnarray}

When $m\neq0$, application of the boundary conditions (\ref{BCsCRVD2}) as described in Section~\ref{BVP2C1} yields six conditions involving the constants $\left\{A_1,A_2,A_3,B_1,B_2,B_3\right\}$ that  can again be cast in the form (\ref{BCmatEQC11}), where, in the present case, the $3\times3$ matrix blocks $\displaystyle\{\mathbf{A}_i,\mathbf{B}_i:i=1,2\}$ are 
\begin{subequations}\label{BCmatEQC2}  
\begin{eqnarray}\label{BCmatEQC21A}
\mathbf{A}_i=\left[\begin{array}{ccc}
            F_{m,i}+V_{m{+}1,i} & G_{m,i}+W_{m{+}1,i} & (m{-}1)Q_{m,i}-mW_{m{+}1,i} \\
            (m{-}1)P_{m,i}-mV_{m{+}1,i} & ~(m{-}1)Q_{m,i}-mW_{m{+}1,i}~ & H_{m,i}+W_{m{+}1,i} \\
            2(P_{m,i}-V_{m{+}1,i}) & (1{+}\gamma_2)(Q_{m,i}-W_{m{+}1,i}) & Q_{m,i}  
            \end{array} \right], ~~
\end{eqnarray}
\begin{eqnarray}\label{BCmatEQC21B}
\mathbf{B}_i=\left[\begin{array}{ccc}
\widetilde{F}_{m,i}+\widetilde{V}_{m{+}1,i} & \widetilde{G}_{m,i}+\widetilde{W}_{m{+}1,i} & (m{-}1)\widetilde{Q}_{m,i}-m\widetilde{W}_{m{+}1,i} \\
(m{-}1)\widetilde{P}_{m,i}-m\widetilde{V}_{m{+}1,i} & ~(m{-}1)\widetilde{Q}_{m,i}-m\widetilde{W}_{m{+}1,i}~ & \widetilde{H}_{m,i}+\widetilde{W}_{m{+}1,i} \\
2\left(\widetilde{P}_{m,i}-\widetilde{V}_{m{+}1,i}\right) & (1{+}\gamma_2)\left(\widetilde{Q}_{m,i}-\widetilde{W}_{m{+}1,i}\right) & \widetilde{Q}_{m,i} \\
\end{array} \right], ~~
\end{eqnarray}
and the $3\times1$ column blocks $\displaystyle\{\mathbf{X}_A,\mathbf{X}_B, \mathbf{S}_1, \mathbf{S}_2\}$ are as given by (\ref{BCmatEQC11C}) and (\ref{BCmatEQC17}). 
The shorthand notation employed in Eqs.~(\ref{BCmatEQC21A}) and (\ref{BCmatEQC21B}) is as follows:
\begin{eqnarray}\label{BCmatEQC22a}
\left\{\begin{array}{c} F_{m,i} \\ \widetilde{F}_{m,i} \end{array} \right\} \equiv \left(-{\eta_1R_i\over2\mu}+{m(m-1)\over R_i}\right) \left\{\begin{array}{c} J_m(\alpha_1R_i) \\ Y_m(\alpha_1R_i) \end{array} \right\}, \quad i=1,2 
\end{eqnarray}
\begin{eqnarray}\label{BCmatEQC22b}
\left\{\begin{array}{c} G_{m,i} \\ \widetilde{G}_{m,i} \end{array} \right\} \equiv \left(-{\eta_2R_i\over2\mu}+{m(m-1)\over R_i}\right) \left\{\begin{array}{c} J_m(\alpha_2R_i) \\ Y_m(\alpha_2R_i) \end{array} \right\}, \quad i=1,2 
\end{eqnarray}
\begin{eqnarray}\label{BCmatEQC22c}
\left\{\begin{array}{c} H_{m,i} \\ \widetilde{H}_{m,i} \end{array} \right\} \equiv \left(-{\alpha^2_2R_i\over2}+{m(m-1)\over R_i}\right) \left\{\begin{array}{c} J_m(\alpha_2R_i) \\ Y_m(\alpha_2R_i) \end{array} \right\}, \quad i=1,2 
\end{eqnarray}
\begin{eqnarray}\label{BCmatEQC23}
\left\{\begin{array}{c} P_{m,i} \\ \widetilde{P}_{m,i} \end{array} \right\} \equiv {m\over R_i} \left\{\begin{array}{c} J_m(\alpha_1R_i) \\ Y_m(\alpha_1R_i) \end{array} \right\}, \quad i=1,2 
\end{eqnarray}
\begin{eqnarray}\label{BCmatEQC24}
\left\{\begin{array}{c} Q_{m,i} \\ \widetilde{Q}_{m,i} \end{array} \right\} \equiv {m\over R_i} \left\{\begin{array}{c} J_m(\alpha_2R_i) \\ Y_m(\alpha_2R_i) \end{array} \right\}, \quad i=1,2 
\end{eqnarray}
\begin{eqnarray}\label{BCmatEQC25}
\left\{\begin{array}{c} V_{m{+}1,i} \\ \widetilde{V}_{m{+}1,i} \end{array} \right\} \equiv \alpha_1 \left\{\begin{array}{c} J_{m+1}(\alpha_1R_i) \\ Y_{m+1}(\alpha_1R_i) \end{array} \right\}, \quad i=1,2 
\end{eqnarray}
\begin{eqnarray}\label{BCmatEQC26}
\left\{\begin{array}{c} W_{m{+}1,i} \\ \widetilde{W}_{m{+}1,i} \end{array} \right\} \equiv \alpha_2 \left\{\begin{array}{c} J_{m+1}(\alpha_2R_i) \\ Y_{m+1}(\alpha_2R_i) \end{array} \right\}, \quad i=1,2. 
\end{eqnarray}
\end{subequations} 

\subsubsection{Special Case: $\displaystyle m=0$}

When $m=0$, $u_\theta=0$, $\sigma_{r\theta}=0$, and boundary conditions (\ref{strssrt2}) are identically satisfied. Application of boundary conditions (\ref{strssrr2}) and (\ref{strssrz2}) yields the $4\times4$ linear system: 
\begin{eqnarray}\label{BCmatEQC2SCm0}
\left[\begin{array}{cccc}
            F_{0,1}+V_{1,1} & G_{0,1}+W_{1,1} & \widetilde{F}_{0,1}+\widetilde{V}_{1,1} & \widetilde{G}_{0,1}+\widetilde{W}_{1,1} \\
            -2V_{1,1} & ~-(1{+}\gamma_2)W_{1,1}~ & -2\widetilde{V}_{1,1} & -(1{+}\gamma_2)\widetilde{W}_{1,1} \\  
            F_{0,2}+V_{1,2} & G_{0,2}+W_{1,2} & ~\widetilde{F}_{0,2}+\widetilde{V}_{1,2}~ & \widetilde{G}_{0,2}+\widetilde{W}_{1,2} \\
            -2V_{1,2} & -(1{+}\gamma_2)W_{1,2} & -2\widetilde{V}_{1,2} & -(1{+}\gamma_2)\widetilde{W}_{1,2} 
           \end{array} \right] \left[\begin{array}{c}
             A_1 \\
             A_2 \\
             B_1 \\ 
             B_2 
           \end{array} \right] =  \left[\begin{array}{c}
              \mathbb{A} \\
              \mathbb{C} \\ 
              \mathbb{D} \\
              \mathbb{F} 
           \end{array} \right],
\end{eqnarray}
where the matrix elements are the evaluated zero- and first-order Bessel functions obtained from substituting $m=0$ in Eqs.~(\ref{BCmatEQC22a}), (\ref{BCmatEQC22b}), (\ref{BCmatEQC25}), and (\ref{BCmatEQC26}). 

In this special case, the non-zero components of the displacement field reduce to:
\begin{subequations}\label{SolBVP2C2meq0}
\begin{equation}\label{SolBVP2C2meq01}
u_r(r,z,t)=-\Bigg\{\sum_{s=1}^2\alpha_s\Big[A_sJ_{1}(\alpha_sr)+B_sY_{1}(\alpha_sr)\Big]\Bigg\}\sin\left({k\pi\over L}z\right)\sin(\omega t), 
\end{equation}
\begin{equation}\label{SolBVP2C2meq02}
u_z(r,z,t)=\left({k\pi\over L}\right)\Bigg\{\sum_{s=1}^2\gamma_s\Big[A_sJ_0(\alpha_sr)+B_sY_0(\alpha_sr)\Big]\Bigg\}\cos\left({k\pi\over L}z\right)\sin(\omega t),   
\end{equation}
\end{subequations}
where constants $\alpha_s$ and $\gamma_s$ are given by Eqs.~(\ref{alphasEG5}) and (\ref{solpsiPP2EG}), respectively, and the constants $\left\{A_1,A_2,B_1,B_2\right\}$ are those obtained from solving Eq.~(\ref{BCmatEQC2SCm0}).  

\subsection{\textbf{Case 3:} $\displaystyle {\rho\omega^2\over(\lambda+2\mu)}<\left({k\pi\over L}\right)^2<{\rho\omega^2\over\mu}$}\label{BVP1C3}

According to Tables \ref{TabLinCombos} and \ref{EG1Cases}, the $s=1$ term in each of the radial parts of Eqs.~(\ref{SolB1})-(\ref{SolB3}) should employ the modified Bessel functions $\left\{I_m(\alpha_1r),K_m(\alpha_1r)\right\}$ (and their derivatives) while the $s=2$ terms should employ the Bessel functions $\left\{J_m(\alpha_2r),Y_m(\alpha_2r)\right\}$ (and their derivatives). The remaining terms are unmodified from those of Case 2. The displacement components thus take the form:
\begin{subequations}\label{GenSolBC3}
\begin{eqnarray}\label{SolBC31}
u_r&=&\Bigg\{A_1\bigg[{m\over r}I_m(\alpha_1r)+\alpha_1I_{m+1}(\alpha_1r)\bigg]+B_1\bigg[{m\over r}K_m(\alpha_1r)-\alpha_1K_{m+1}(\alpha_1r)\bigg] \nonumber \\ 
&&+~A_2\bigg[{m\over r}J_m(\alpha_2r)-\alpha_2J_{m+1}(\alpha_2r)\bigg] +B_2\bigg[{m\over r}Y_m(\alpha_2r)-\alpha_2Y_{m+1}(\alpha_2r)\bigg] \nonumber \\
&&+~{m\over r}\Big[A_3J_m(\alpha_2r)+B_3Y_m(\alpha_2r)\Big]\Bigg\}\cos(m\theta)\sin\left({k\pi\over L}z\right)\sin(\omega t),   
\end{eqnarray}
\begin{eqnarray}\label{SolBC32}
u_\theta&=&-\Bigg\{{m\over r}\Big[A_1I_m(\alpha_1r)+B_1K_m(\alpha_1r)+A_2J_m(\alpha_2r)+B_2Y_m(\alpha_2r)\Big] \nonumber \\
&&~~~~+~A_3\bigg[{m\over r}J_m(\alpha_2r)-\alpha_2J_{m+1}(\alpha_2r)\bigg]+B_3\bigg[{m\over r}Y_m(\alpha_2r)-\alpha_2Y_{m+1}(\alpha_2r)\bigg]\Bigg\} \nonumber \\
&& \quad \quad \quad \quad \times \sin(m\theta)\sin\left({k\pi\over L}z\right)\sin(\omega t), 
\end{eqnarray}
\begin{eqnarray}\label{SolBC33}
u_z&=&\left({k\pi\over L}\right)\Bigg\{\gamma_1\Big[A_1I_m(\alpha_1r)+B_1K_m(\alpha_1r)\Big]+\gamma_2\Big[A_2J_m(\alpha_2r)+B_2Y_m(\alpha_2r)\Big]\Bigg\} \nonumber \\
&& \quad \quad \quad \quad \times \cos(m\theta)\cos\left({k\pi\over L}z\right)\sin(\omega t),  
\end{eqnarray}
\end{subequations}
where 
\begin{equation}\label{alphasEG3}
\alpha_1=\sqrt{\left({k\pi\over L}\right)^2-{\rho\omega^2\over(\lambda+2\mu)}}, \quad \alpha_2=\sqrt{-\left({k\pi\over L}\right)^2+{\rho\omega^2\over\mu}},
\end{equation}
and $\gamma_s$ is again given by Eq.~(\ref{solpsiPP2EG}).

The constants $\left\{A_1,A_2,A_3,B_1,B_2,B_3\right\}$ in Eqs.~(\ref{SolBC31})-(\ref{SolBC33}) must again be chosen so as to satisfy boundary conditions (\ref{BCsCRVD2}). Proceeding as usual, we first obtain the pertinent components of the stress field. Substituting Eqs.~(\ref{SolBC31})-(\ref{SolBC33}) into Eqs.~(\ref{stssstrnCYL1}), (\ref{stssstrnCYL4}), and (\ref{stssstrnCYL5}), and performing the  necessary algebra yields the required stress components:
\begin{eqnarray}\label{strssRRC3}
\sigma_{rr}(r,\theta,z,t)&=&2\mu\Bigg\{A_1\left[\left({\beta_1\over2\mu}+{m(m-1)\over r^2}\right)I_m(\alpha_1r)-{\alpha_1\over r}I_{m+1}(\alpha_1r)\right] \nonumber \\
&&+~A_2\left[\left(-{\eta_2\over2\mu}+{m(m-1)\over r^2}\right)J_m(\alpha_2r)+{\alpha_2\over r}J_{m+1}(\alpha_2r)\right] \nonumber \\ 
&&+~A_3\left[{m(m-1)\over r^2}J_m(\alpha_2r)-{\alpha_2m\over r}J_{m+1}(\alpha_2r)\right] \nonumber \\
&&+~B_1\left[\left({\beta_1\over2\mu}+{m(m-1)\over r^2}\right)K_m(\alpha_1r)+{\alpha_1\over r}K_{m+1}(\alpha_1r)\right] \nonumber \\ 
&&+~B_2\left[\left(-{\eta_2\over2\mu}+{m(m-1)\over r^2}\right)Y_m(\alpha_2r)+{\alpha_2\over r}Y_{m+1}(\alpha_2r)\right] \nonumber \\
&&+~B_3\left[{m(m-1)\over r^2}Y_m(\alpha_2r)-{\alpha_2m\over r}Y_{m+1}(\alpha_2r)\right] \Bigg\} \nonumber \\ 
&& \quad \quad \times \cos(m\theta)\sin\left({k\pi\over L}z\right)\sin(\omega t),
\end{eqnarray}
where the constants $\beta_1$ and $\eta_2$ are as given by Eqs.~(\ref{strsscmpRRC11B}) and (\ref{strsscmpRRC2P1B}), respectively,
\begin{eqnarray}\label{strssRTC3}
\sigma_{r\theta}(r,\theta,z,t)&=&-2\mu\Bigg\{A_1\bigg[{m(m-1)\over r^2}I_m(\alpha_1r)+{\alpha_1m\over r}I_{m+1}(\alpha_1r)\bigg] \nonumber \\
&&+~A_2\bigg[{m(m-1)\over r^2}J_m(\alpha_2r)-{\alpha_2m\over r}J_{m+1}(\alpha_2r)\bigg] \nonumber \\ 
&&+~A_3\left[\bigg({m(m-1)\over r^2}-{\alpha_2^2\over2}\bigg)J_m(\alpha_2r)+{\alpha_2\over r}J_{m+1}(\alpha_2r)\right] \nonumber \\
&&+~B_1\bigg[{m(m-1)\over r^2}K_m(\alpha_1r)-{\alpha_1m\over r}K_{m+1}(\alpha_1r)\bigg] \nonumber \\ 
&&+~B_2\bigg[{m(m-1)\over r^2}Y_m(\alpha_2r)-{\alpha_2m\over r}Y_{m+1}(\alpha_2r)\bigg] \nonumber \\ 
&&+~B_3\left[\bigg({m(m-1)\over r^2}-{\alpha_2^2\over2}\bigg)Y_m(\alpha_2r)+{\alpha_2\over r}Y_{m+1}(\alpha_2r)\right] \Bigg\} \nonumber \\
&& \quad \quad \quad \times \sin(m\theta)\sin\left({k\pi\over L}z\right)\sin(\omega t),
\end{eqnarray}
and 
\begin{eqnarray}\label{strssRZC3}
\sigma_{rz}(r,\theta,z,t)&=&\mu\left({k\pi\over L}\right)\Bigg\{A_1\left(1+\gamma_1\right)\bigg[{m\over r}I_m(\alpha_1r)+\alpha_1I_{m+1}(\alpha_1r)\bigg] \nonumber \\ 
&+&~A_2\left(1+\gamma_2\right)\bigg[{m\over r}J_m(\alpha_2r)-\alpha_2J_{m+1}(\alpha_2r)\bigg] + A_3\bigg[{m\over r}J_m(\alpha_2r)\bigg] \nonumber \\ 
&+&~B_1\left(1+\gamma_1\right)\bigg[{m\over r}K_m(\alpha_1r)-\alpha_1K_{m+1}(\alpha_1r)\bigg] \nonumber \\ 
&+&~B_2\left(1+\gamma_2\right)\bigg[{m\over r}Y_m(\alpha_2r)-\alpha_2Y_{m+1}(\alpha_2r)\bigg] \nonumber \\ 
&+&~B_3\bigg[{m\over r}Y_m(\alpha_2r)\bigg]\Bigg\}\cos(m\theta)\cos\left({k\pi\over L}z\right)\sin(\omega t).
\end{eqnarray} 

When $m\neq0$, application of the boundary conditions as described in Section~\ref{BVP2C1} yields a $6\times6$ linear system that can again be cast in the form (\ref{BCmatEQC11}); in the present case, the $3\times3$ matrix blocks $\displaystyle\{\mathbf{A}_i,\mathbf{B}_i:i=1,2\}$ are
\begin{subequations}\label{BCmatEQC3} 
\begin{eqnarray}\label{BCmatEQC31A}
\mathbf{A}_i=\left[\begin{array}{ccc}
            f_{m,i}-v_{m{+}1,i} & G_{m,i}+W_{m{+}1,i} & (m{-}1)Q_{m,i}-mW_{m{+}1,i} \\
            (m{-}1)p_{m,i}+mv_{m{+}1,i} & ~(m{-}1)Q_{m,i}-mW_{m{+}1,i}~ & H_{m,i}+W_{m{+}1,i} \\
            2(p_{m,i}+v_{m{+}1,i}) & (1{+}\gamma_2)(Q_{m,i}-W_{m{+}1,i}) & Q_{m,i}  
            \end{array} \right], ~~
\end{eqnarray}
\begin{eqnarray}\label{BCmatEQC31B}
\mathbf{B}_i=\left[\begin{array}{ccc}
\widetilde{f}_{m,i}+\widetilde{v}_{m{+}1,i} & \widetilde{G}_{m,i}+\widetilde{W}_{m{+}1,i} & (m{-}1)\widetilde{Q}_{m,i}-m\widetilde{W}_{m{+}1,i} \\
(m{-}1)\widetilde{p}_{m,i}-m\widetilde{v}_{m{+}1,i} & ~(m{-}1)\widetilde{Q}_{m,i}-m\widetilde{W}_{m{+}1,i}~ & \widetilde{H}_{m,i}+\widetilde{W}_{m{+}1,i} \\
2\left(\widetilde{p}_{m,i}-\widetilde{v}_{m{+}1,i}\right) & (1{+}\gamma_2)\left(\widetilde{Q}_{m,i}-\widetilde{W}_{m{+}1,i}\right) & \widetilde{Q}_{m,i} \\
\end{array} \right], ~~
\end{eqnarray} 
\end{subequations}
and the $3\times1$ column blocks $\displaystyle\{\mathbf{X}_A,\mathbf{X}_B, \mathbf{S}_1, \mathbf{S}_2\}$ are as given by (\ref{BCmatEQC11C}) and (\ref{BCmatEQC17}). 
The shorthand notation employed for all matrix elements in Eqs.~(\ref{BCmatEQC31A}) and (\ref{BCmatEQC31B}) is as previously defined by Eqs.~(\ref{BCmatEQC12a}), (\ref{BCmatEQC13}), (\ref{BCmatEQC15}) and Eqs.~(\ref{BCmatEQC22b}), (\ref{BCmatEQC22c}), (\ref{BCmatEQC24}), (\ref{BCmatEQC26}). 

\subsubsection{Special Case: $\displaystyle m=0$}

When $m=0$, $u_\theta=0$, $\sigma_{r\theta}=0$, and boundary conditions (\ref{strssrt2}) are identically satisfied. Application of boundary conditions (\ref{strssrr2}) and (\ref{strssrz2}) yields the $4\times4$ linear system: 
\begin{eqnarray}\label{BCmatEQC3SCm0}
\left[\begin{array}{cccc}
            f_{0,1}-v_{1,1} & G_{0,1}+W_{1,1} & \widetilde{f}_{0,1}+\widetilde{v}_{1,1} & \widetilde{G}_{0,1}+\widetilde{W}_{1,1} \\
            2v_{1,1} & ~-(1{+}\gamma_2)W_{1,1}~ & -2\widetilde{v}_{1,1} & -(1{+}\gamma_2)\widetilde{W}_{1,1} \\  
            f_{0,2}-v_{1,2} & G_{0,2}+W_{1,2} & ~\widetilde{f}_{0,2}+\widetilde{v}_{1,2}~ & \widetilde{G}_{0,2}+\widetilde{W}_{1,2} \\
            2v_{1,2} & -(1{+}\gamma_2)W_{1,2} & -2\widetilde{v}_{1,2} & -(1{+}\gamma_2)\widetilde{W}_{1,2} 
           \end{array} \right] \left[\begin{array}{c}
             A_1 \\
             A_2 \\
             B_1 \\ 
             B_2 
           \end{array} \right] =  \left[\begin{array}{c}
              \mathbb{A} \\
              \mathbb{C} \\ 
              \mathbb{D} \\
              \mathbb{F} 
           \end{array} \right].
\end{eqnarray}
The matrix elements in Eq.~(\ref{BCmatEQC3SCm0}) are the evaluated zero- and first-order Bessel functions obtained from substituting $m=0$ in Eqs.~(\ref{BCmatEQC12a}), (\ref{BCmatEQC13}), (\ref{BCmatEQC15}) and Eqs.~(\ref{BCmatEQC22b}), (\ref{BCmatEQC22c}), (\ref{BCmatEQC24}), (\ref{BCmatEQC26}). 

In this special case, the non-zero components of the displacement field reduce to:
\begin{subequations}\label{SolBVP2C3meq0}
\begin{equation}\label{SolBVP2C3meq01}
u_r=\Bigg\{A_1\alpha_1I_{1}(\alpha_1r)-B_1\alpha_1K_{1}(\alpha_1r)-A_2\alpha_2J_{1}(\alpha_2r)-B_2\alpha_2Y_{1}(\alpha_2r)\Bigg\}\sin\left({k\pi\over L}z\right)\sin(\omega t), 
\end{equation}
\begin{equation}\label{SolBVP2C3meq02}
u_z=\left({k\pi\over L}\right)\Bigg\{\gamma_1\Big[A_1I_0(\alpha_1r)+B_1K_0(\alpha_1r)\Big]+\gamma_2\Big[A_2J_0(\alpha_2r)+B_2Y_0(\alpha_2r)\Big]\Bigg\}\cos\left({k\pi\over L}z\right)\sin(\omega t), 
\end{equation}
\end{subequations}
where constants $\alpha_s$ and $\gamma_s$ ($s=1,2$) are given by Eqs.~(\ref{alphasEG3}) and (\ref{solpsiPP2EG}), respectively, and the constants $\left\{A_1,A_2,B_1,B_2\right\}$ are those obtained from solving Eq.~(\ref{BCmatEQC3SCm0}). 

\section{Analytics II: Special Case $\displaystyle k=0$}\label{solntoBVPKeq0}

As discussed in Section~\ref{generalKeq0}, a general displacement field compatible with the boundary-value problem defined in Section~\ref{EGFULL} in the special case $k=0$ is given by particular solution (\ref{gensolnkeq0}). We need now only to determine the values of the constants $\displaystyle\left\{A,B\right\}$ in (\ref{gensolnkeq0}) that satisfy boundary conditions (\ref{strssrr2})-(\ref{strssrz2}). Substituting the components of solution (\ref{gensolnkeq0}) into Eqs.~(\ref{stssstrnCYL1}), (\ref{stssstrnCYL4}), and (\ref{stssstrnCYL5}) immediately yields the stress components:
\begin{subequations}\label{strsssolnkeq0}
\begin{equation}\label{strsssolnkeq0A}
\sigma_{rr}(r,\theta,z,t)=\sigma_{r\theta}(r,\theta,z,t)=0,
\end{equation}
and
\begin{equation}\label{strsssolnkeq0B}
\sigma_{rz}(r,\theta,t)=\mu\Bigg\{A\bigg[{m\over r}J_m(\alpha r)-\alpha J_{m+1}(\alpha r)\bigg] + B\bigg[{m\over r}Y_m(\alpha r)-\alpha Y_{m+1}(\alpha r)\bigg]\Bigg\}\cos(m\theta)\sin(\omega t).
\end{equation}
\end{subequations}
Boundary conditions (\ref{strssrr2}) and (\ref{strssrt2}) are therefore satisfied identically. Application of boundary conditions (\ref{strssrz2}) yields two conditions that can be compactly written as the following $2\times2$ linear system: 
\begin{subequations}\label{BCmatEQkeq0}
\begin{eqnarray}\label{BCmatEQkeq0A}
\left[\begin{array}{cc}
            Q_{m,1}-W_{m+1,1} & ~\widetilde{Q}_{m,1}-\widetilde{W}_{m+1,1} \\
            Q_{m,2}-W_{m+1,2} & ~\widetilde{Q}_{m,2}-\widetilde{W}_{m+1,2}
           \end{array} \right] \left[\begin{array}{c}
             A \\ 
             B 
           \end{array} \right] =  \left[\begin{array}{c}
              \mathbb{C} \\ 
              \mathbb{F} 
           \end{array} \right],
\end{eqnarray}
where the following shorthand notation is employed for the matrix elements of Eq.~(\ref{BCmatEQkeq0A}):
\begin{eqnarray}\label{BCmatEQkeq0B}
\left\{\begin{array}{c} Q_{m,i} \\ \widetilde{Q}_{m,i} \end{array} \right\} \equiv {m\over R_i} \left\{\begin{array}{c} J_m(\alpha R_i) \\ Y_m(\alpha R_i) \end{array} \right\}, \quad i=1,2 
\end{eqnarray}
\begin{eqnarray}\label{BCmatEQCkeq0C}
\left\{\begin{array}{c} W_{m{+}1,i} \\ \widetilde{W}_{m{+}1,i} \end{array} \right\} \equiv \alpha \left\{\begin{array}{c} J_{m+1}(\alpha R_i) \\ Y_{m+1}(\alpha R_i) \end{array} \right\}, \quad i=1,2 
\end{eqnarray}
\begin{equation}\label{BCmatEQCkeq0D}
\mathbb{C}\equiv{\mathcal{C}\over\mu}, \quad \mathbb{F}\equiv{\mathcal{F}\over\mu}.
\end{equation}
\end{subequations} 
The general solution to system (\ref{BCmatEQkeq0}) is:
\begin{subequations}\label{solnBcmatEQkeq0}
\begin{equation}\label{solnBcmatEQkeq0A}
A={\left(\widetilde{Q}_{m,2}-\widetilde{W}_{m+1,2}\right)\mathbb{C}-\left(\widetilde{Q}_{m,1}-\widetilde{W}_{m+1,1}\right)\mathbb{F}\over\big(Q_{m,1}-W_{m+1,1}\big)\left(\widetilde{Q}_{m,2}-\widetilde{W}_{m+1,2}\right)-\left(\widetilde{Q}_{m,1}-\widetilde{W}_{m+1,1}\right)\big(Q_{m,2}-W_{m+1,2}\big)},
\end{equation}
\begin{equation}\label{solnBcmatEQkeq0B}
B={\big(Q_{m,1}-W_{m+1,1}\big)\mathbb{F}-\big(Q_{m,2}-W_{m+1,2}\big)\mathbb{C}\over\big(Q_{m,1}-W_{m+1,1}\big)\left(\widetilde{Q}_{m,2}-\widetilde{W}_{m+1,2}\right)-\left(\widetilde{Q}_{m,1}-\widetilde{W}_{m+1,1}\right)\big(Q_{m,2}-W_{m+1,2}\big)}.
\end{equation}
\end{subequations}
When $m=0$, (\ref{solnBcmatEQkeq0}) reduces to:
\begin{subequations}\label{solnBcmatEQkeq0meq0}
\begin{equation}\label{solnBcmatEQkeq0Ameq0}
A={Y_1(\alpha R_1)\mathbb{F}-Y_1(\alpha R_2)\mathbb{C}\over\alpha\Big[J_1(\alpha R_1)Y_1(\alpha R_2)-J_1(\alpha R_2)Y_1(\alpha R_1)\Big]},
\end{equation}
\begin{equation}\label{solnBcmatEQkeq0Bmeq0}
B={J_1(\alpha R_2)\mathbb{C}-J_1(\alpha R_1)\mathbb{F}\over\alpha\Big[J_1(\alpha R_1)Y_1(\alpha R_2)-J_1(\alpha R_2)Y_1(\alpha R_1)\Big]}.
\end{equation}
\end{subequations}
Thus, in the special case $k=0$, the displacement field is given by (\ref{gensolnkeq0}) with the constants $\left\{A,B\right\}$ given by (\ref{solnBcmatEQkeq0}), which reduces to  (\ref{solnBcmatEQkeq0meq0}) when $m=0$. 

\section{Consistency with the ERP Field Equations}\label{consisSOLNs}

As an analytical check, we have verified that, in the special $m=0$ case, the general stress and displacement fields obtained from applying our method of solution agree with the general axisymmetric field equations that would be obtained from applying the mathematical framework of Ebenezer et al. (ERP) \cite{Ebenezer15}. Demonstration of this consistency is somewhat intricate; the details are therefore consigned to Appendix \ref{appendx2}. Equivalency of our $m=0$ solution with the axisymmetric solution that would be obtained from the ERP method then directly follows from application of the boundary conditions. 

\section{Example 1}\label{numerica}

\begin{table}[h]
\centering
\begin{tabular}{ll}
    \hline  \hline
     Parameter & Numerical Value \\ \hline
     Length ($L$) & $0.300$ m \\ 
     Outer radius ($R_2$) & $0.100$ m \\
     Inner radius ($R_1$) & $0.050$ m  \\ 
     Mass density ($\rho$) & $8000~\text{kg/m}^3$  \\ 
     Young modulus ($E$)  & $190$ GPa \\ 
     Poisson ratio ($\nu$)  & $0.300$  \\ 
     First Lam\'{e} constant ($\lambda$)  & $110$ GPa \\ 
     Second Lam\'{e} constant ($\mu$) \hspace*{2cm} & $73.1$ GPa \\ \hline  \hline 
\end{tabular}
\caption{Geometric and material parameter values used in Example 1.}
\label{GeoMatParams}
\end{table}

\begin{table}[h]
\centering 
\begin{tabular}{ccccccccc}
    \hline  \hline 
    Mode number &~~~~~& \multicolumn{7}{c}{Circumferential wave number} \\ 
    \cline{3-9} 
    $n$ && $m=0$ &~~~~~& $m=1$ &~~~~~& $m=2$ &~~~~~& $m=3$ \\ \hline
    1 && 7.698 &~~~~~& 2.997 &~~~~~& 5.391 &~~~~~& 11.537 \\
    2 && 10.686 &~~~~~& 6.515 &~~~~~& 8.141 &~~~~~& 13.117  \\
    3 && 11.827 &~~~~~& 7.284 &~~~~~& 12.154 &~~~~~& 15.954 \\
    4 && 12.838 &~~~~~& 9.731 &~~~~~& 12.896 &~~~~~& 19.035 \\
    5 && 16.229 &~~~~~& 11.486 &~~~~~& 14.429 &~~~~~& 19.641 \\
    6 && 17.542 &~~~~~& 13.563 &~~~~~& 16.625  &~~~~~& 19.964 \\
    7 && 20.398 &~~~~~& 15.467 &~~~~~& 17.460 &~~~~~& 22.222 \\
    8 && 24.963 &~~~~~& 15.828 &~~~~~& 20.912 &~~~~~& 23.805 \\
    9 && 25.425 &~~~~~& 17.334 &~~~~~& 21.290 &~~~~~& 25.163 \\ \hline  \hline    
\end{tabular}
\caption{Natural frequencies $\left\{f^{(m)}_{_{\scriptstyle n}}:n=1,\ldots,9\right\}$ of a simply-supported isotropic (thick-walled) hollow elastic circular cylinder having the geometrical and material properties given in Table \ref{GeoMatParams}. All values are in units of kHz. The above frequencies were computed using the free-vibration frequency data given in Table 6 of Ref.~\cite{Ye14}.}
\label{TabNatFreqs1}
\end{table}

\begin{figure}[h]
\vspace*{-0.5cm}
\centering
\scalebox{0.6}{\includegraphics*{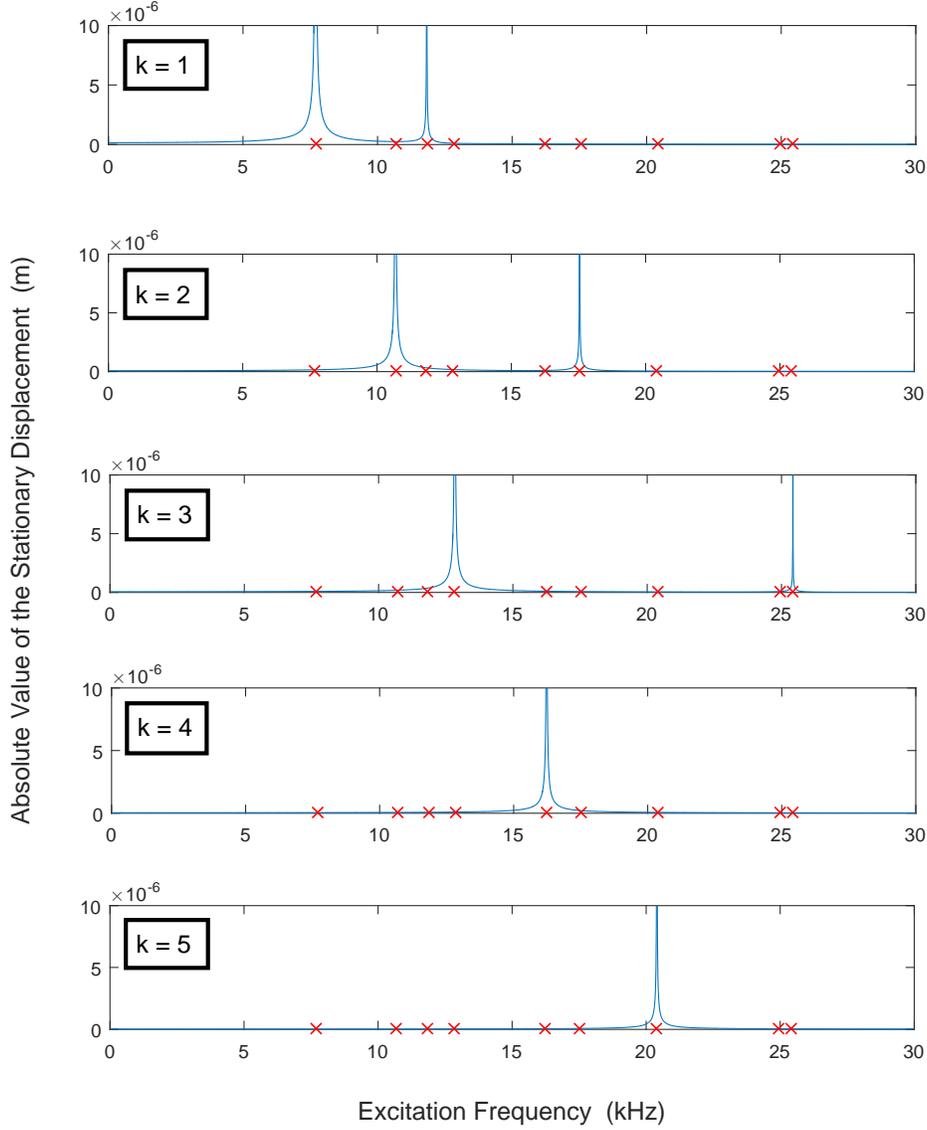}}
%\vspace*{-1cm}
\caption{\label{freqresmeq0} Frequency response of the displacement at the point $(r,\theta,z)=((R_1+R_2)/2,\pi/5,L/7)$ for various values of the longitudinal wave number $k$ and circumferential wave number $m=0$. For reference, the natural frequencies listed in Table \ref{TabNatFreqs1} are marked by `$\mathsf{X}$'s on the frequency axis of each subplot.}
\end{figure}

\begin{figure}[h]
\vspace*{-0.5cm}
\centering
\scalebox{0.6}{\includegraphics*{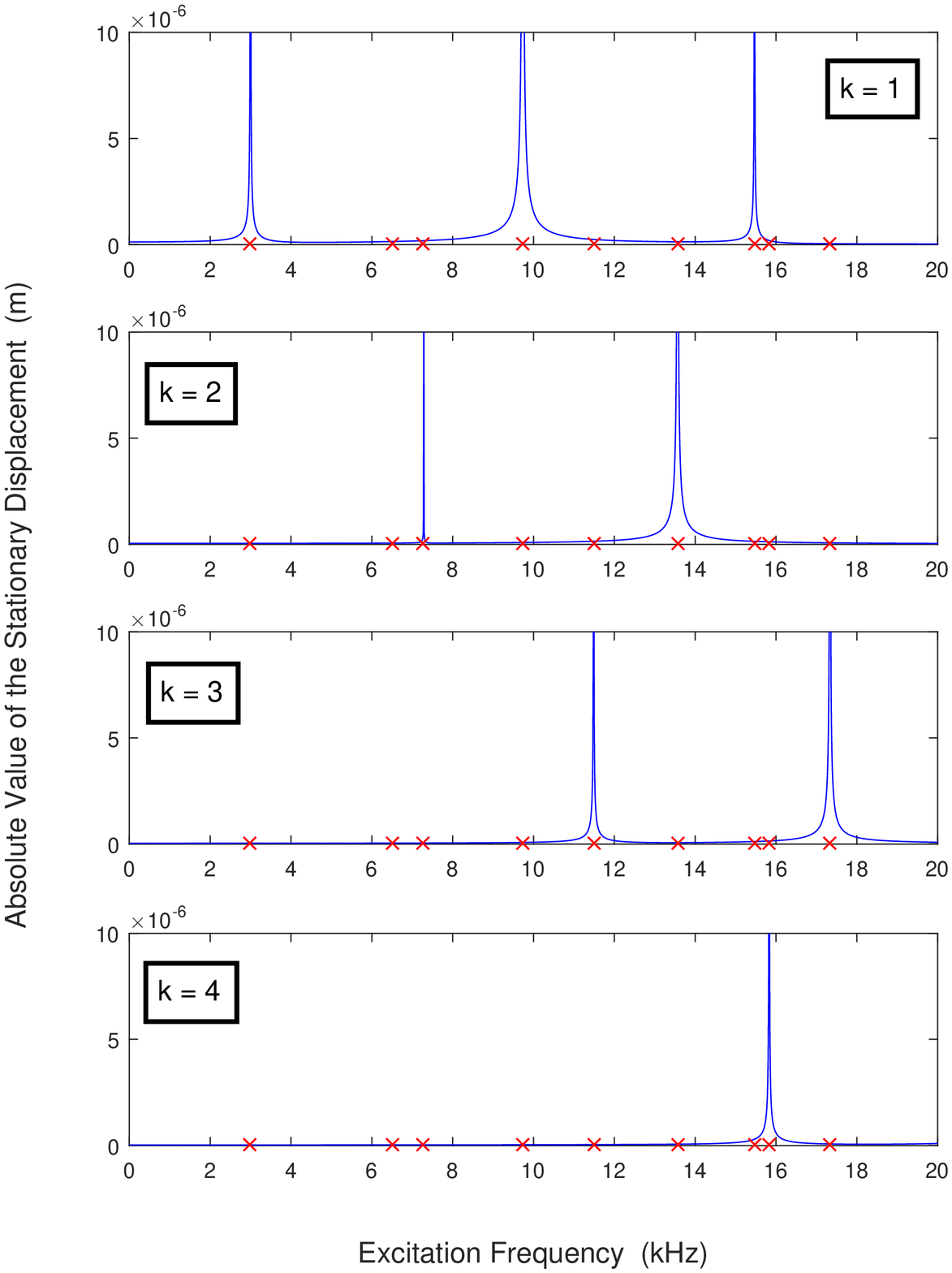}}
%\vspace*{-1cm}
\caption{\label{freqresmeq1} Same as Figure \ref{freqresmeq0} except the circumferential wave number $m=1$.}
\end{figure}

\begin{figure}[h]
\vspace*{-0.5cm}
\centering
\scalebox{0.63}{\includegraphics*{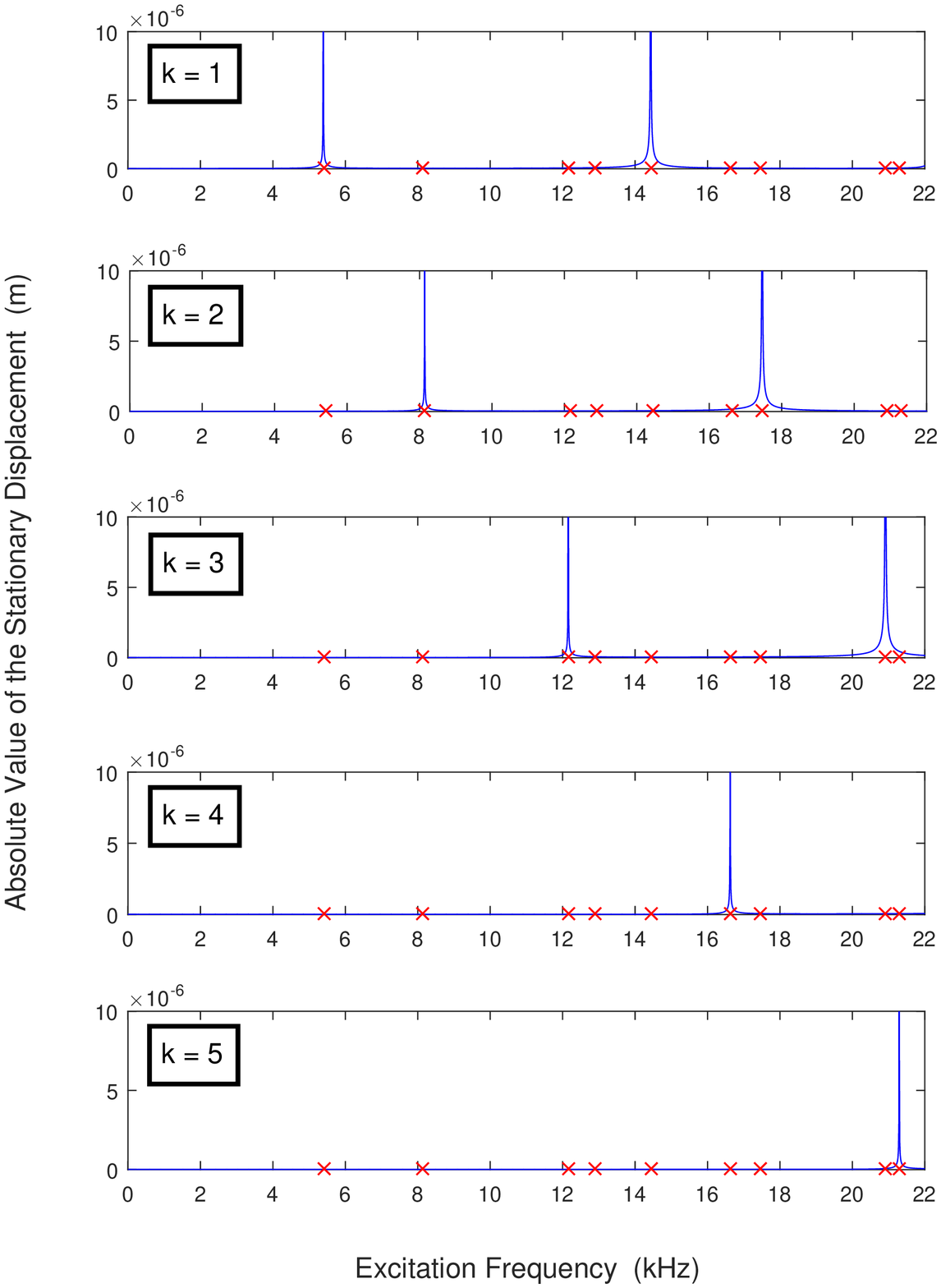}}
%\vspace*{-1cm}
\caption{\label{freqresmeq2} Same as Figure \ref{freqresmeq0} except the circumferential wave number $m=2$.}
\end{figure}

\begin{figure}[h]
\vspace*{-0.5cm}
\centering
\scalebox{0.63}{\includegraphics*{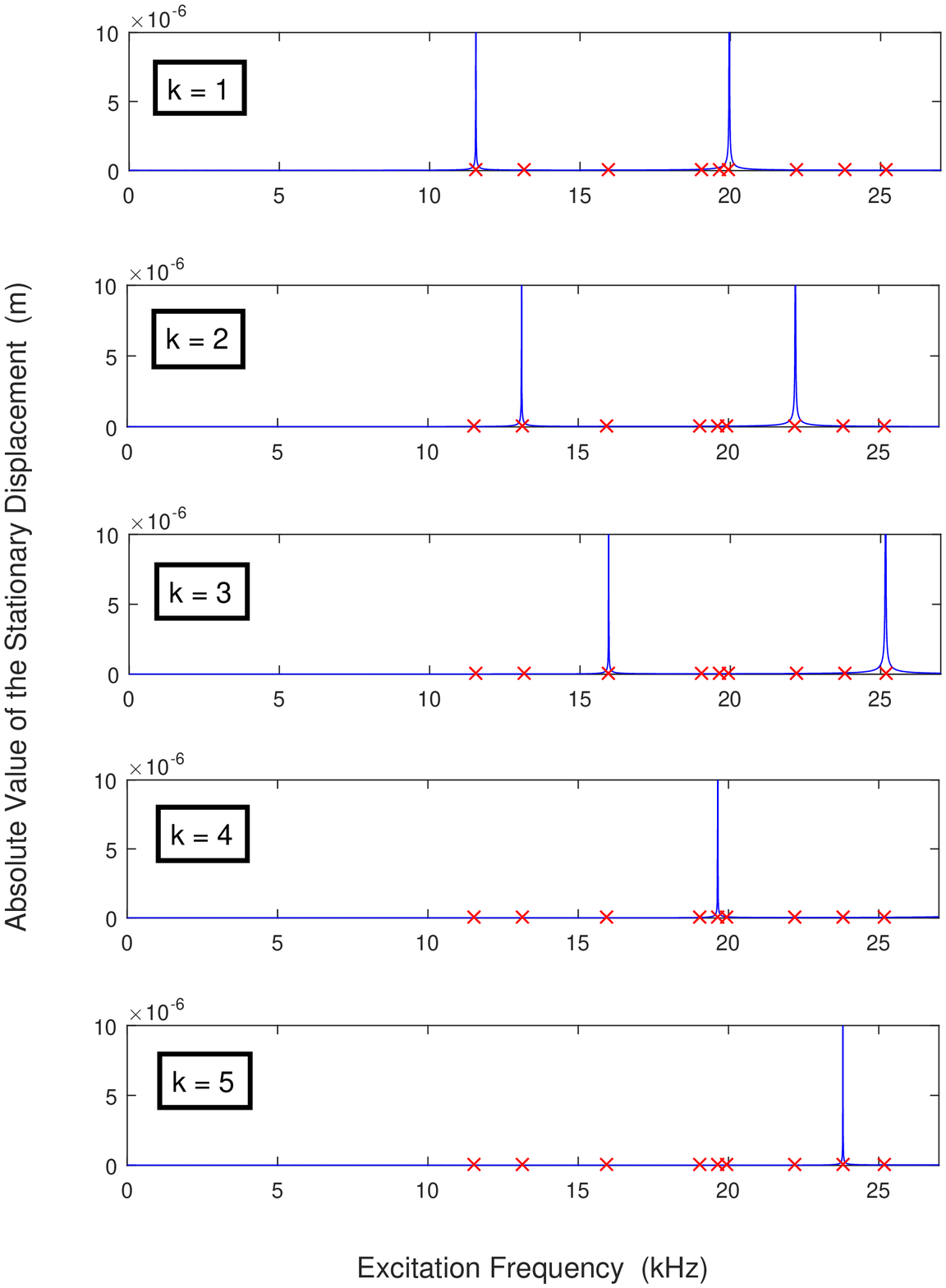}}
%\vspace*{-1cm}
\caption{\label{freqresmeq3} Same as Figure \ref{freqresmeq0} except the circumferential wave number $m=3$.}
\end{figure} 

\begin{figure}[h]
\vspace*{-1cm}
\centering
\scalebox{0.6}{\includegraphics*{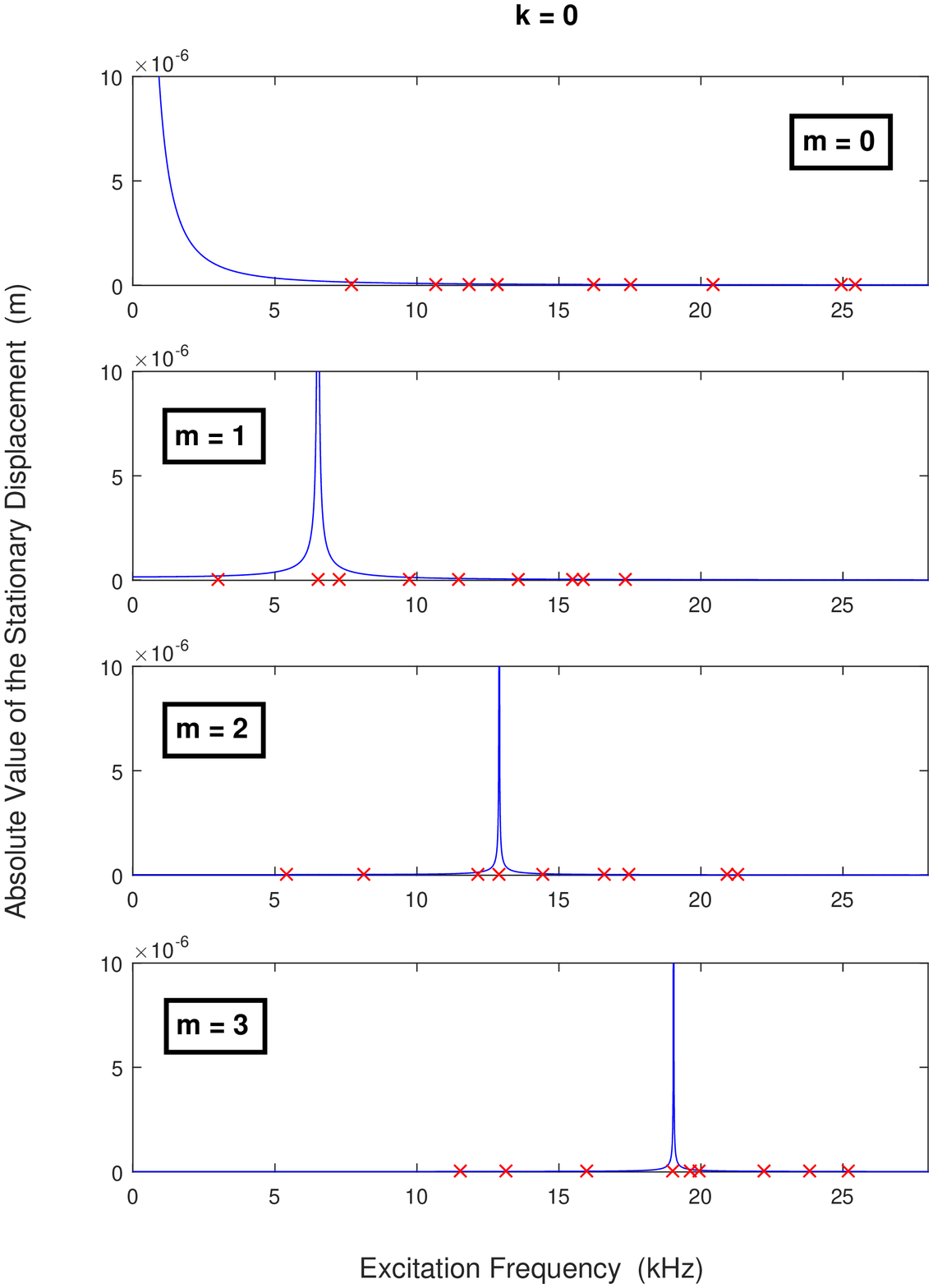}}
%\vspace*{-1cm}
\caption{\label{freqreskeq0} Frequency response of the displacement at the point $(r,\theta,z)=((R_1+R_2)/2,\pi/5,L/7)$ for various values of the circumferential wave number $m$ and longitudinal wave number $k=0$.}
\end{figure} 

As a numerical example, we examine the steady-state frequency response of a thick-walled steel cylinder whose geometric and material properties are specified in Table \ref{GeoMatParams}. As in other  steady-state frequency-response analyses (see, for example, Ref.~\cite{Ebenezer15}), we shall here restrict attention to studying the behavior of the stationary displacement field at a few suitably-chosen representative points in the cylinder as a function of the excitation frequency.\footnote{Aside from nodal or semi-nodal points, we are free to choose any point in the cylinder as a representative point.} To do so, we numerically evaluate the formulas for the displacement field obtained in Section~\ref{solntoBVP}. The only non-trivial numerical detail is the determination of the (frequency-dependent) solution constants $\left\{A_s,B_s: s=1,2,3\right\}$, which we obtain by numerically solving the linear systems (\ref{BCmatEQC1}), (\ref{BCmatEQC2}), and (\ref{BCmatEQC3}) (or their $m=0$ analogs) pointwise for each excitation frequency. An alternative is to use a symbolic algebra package, solve these linear systems symbolically, and then evaluate the results at the excitation frequencies of interest. While it is possible to obtain exact analytical expressions for each of the solution constants (using a symbolic algebra package or otherwise), the resulting expressions are too algebraically complicated for general use. Note that no numerical solution is required when the longitudinal wave number $k=0$ since the equivalent solution constants $\left\{A,B\right\}$ in this special case are given, in closed form, by (\ref{solnBcmatEQkeq0}). 

Using the parameter values given in Table \ref{GeoMatParams}, the components of the displacement field were computed at excitation frequencies that are integer multiples of 1 Hz with lower and upper bounds of 1 Hz and 50 kHz, respectively. In all calculations, the excitation amplitudes were set as follows: $\mathcal{A}=\mathcal{B}=\mathcal{C}=10^5$ Pa and $\mathcal{D}=\mathcal{E}=\mathcal{F}=-{10^5\over2}$ Pa. Given that the cylinder is being forced to vibrate, we expect to observe large displacements (i.e., resonances) when the excitation frequency is close to one of the natural frequencies of the \emph{simply-supported} cylinder. For circumferential wave numbers $m=\{0,1,2,3\}$, the first nine of these frequencies $\left\{f^{(m)}_{_{\scriptstyle n}}:n=1,\ldots,9\right\}$, computed using free-vibration frequency data from Ref.~\cite{Ye14}, are given in Table \ref{TabNatFreqs1}. The absolute value of the (stationary) displacement at the interior point $(r,\theta,z)=((R_1+R_2)/2,\pi/5,L/7)$ is shown in Figs.~\ref{freqresmeq0}-\ref{freqresmeq3} for various values of the longitudinal wave number $k$ and circumferential wave numbers $m=0$, $1$, $2$, and $3$, respectively. For visual reference, the natural frequencies listed in Table \ref{TabNatFreqs1} are marked by `$\mathsf{X}$'s on the frequency axis of each subplot.

In each case, we observe unmistakable resonances around (a subset of) the natural frequencies of the simply-supported cylinder. While the results may at first appear to be particularly simple, there are several interesting features that should be noted. First, note that each  \emph{individual} excitation (obtained by specifying a single pair of $(m,k)$ values) generates a unique \emph{series} of resonances, as opposed to producing just one resonance. In other words, a single harmonic excitation excites a set of resonant modes instead of exciting only one resonant mode. Unfortunately, there does not appear to be any mathematical rule for predicting which resonances will be excited by a particular excitation. More precisely, if (for a given circumferential wave number $m$) $\left\{f^{(m)}_{_{\scriptstyle n}}:n\in\mathbb{Z}^+\right\}$ denotes the complete natural frequency spectrum and $\left\{f^{(m)}_j:j\in S(m,k)\subset\mathbb{Z}^+\right\}$ denotes the subset of the natural frequency spectrum at which an excitation with wave numbers $(m,k)$ generates resonances, then there appears to be no deterministic rule for predicting the set of mode numbers $\{j:j\in S(m,k)\subset\mathbb{Z}^+\}$ given the wave numbers $(m,k)$ of the excitation. Second, note that the resonances generated by any single excitation generally have different widths; in other words, the resonant modes excited by a particular excitation generally possess different decay properties. Practically speaking, this means that the displacement response to any harmonic excitation will have a varying degree of significance in the neighborhoods of the associated resonant frequencies $\left\{f^{(m)}_j\right\}$. For example, when the circumferential wave number $m=1$, the response to a standing-wave excitation with longitudinal wave number $k=1$ is significant at more frequencies neighboring $f^{(1)}_{_{\scriptstyle 4}}=9.731$ kHz than neighboring $f^{(1)}_{_{\scriptstyle 1}}=2.997$ kHz or $f^{(1)}_{_{\scriptstyle 7}}=15.467$ kHz. 

One other noteworthy feature in Figs.~\ref{freqresmeq0}-\ref{freqresmeq3} is the conspicuous absence of resonances in the neighborhoods of certain natural frequencies, in particular, around $f^{(0)}_8=24.963$ kHz (when $m=0$), around $f^{(1)}_2=6.515$ kHz (when $m=1$), around $f^{(2)}_4=12.896$ kHz (when $m=2$), and around $f^{(3)}_4=19.035$ kHz (when $m=3$). As it turns out, when $m\neq0$, resonances at these frequencies are produced by boundary stresses of type (\ref{BCsCRVD2}) with longitudinal wave number $k=0$, as shown in Fig.~\ref{freqreskeq0}. The same figure also shows that, when $m=0$, no resonance associated with $f^{(0)}_8=24.963$ kHz is produced by such an excitation. Thus, when $m=0$, there exist resonant modes at certain frequencies that cannot be excited by harmonic boundary stresses of type (\ref{BCsCRVD2}). 

In Figs.~\ref{freqresmeq0}-\ref{freqreskeq0}, we used common vertical scales in all subplots in order to make it easier to compare the different cases. We should however mention that the amplitudes of the resonances are not all equal, and this is evident when one views the displacement response outside the common vertical range shown in the figures. Differences in amplitude not only occur between the different excitation cases; the amplitudes of the resonances generated by each individual excitation also vary. While it may be obvious to some readers, it is worth emphasizing that the amplitudes, which inherently depend on the frequency resolution\footnote{As previously stated, the components of the displacement field were computed at excitation frequencies that are integer multiples of 1 Hz with lower and upper bounds of 1 Hz and 50 kHz, respectively. When a different frequency discretization is used (e.g., 2 Hz instead of 1 Hz), the numerical amplitudes change.} and on where in the cylinder the displacement is evaluated, should not be interpreted as resonance intensities. In the present context of a lossless (i.e., undamped) cylinder, the amplitudes are insignificant since, in theory, the amplitude of any resonance asymptotically approaches infinity as the excitation frequency approaches the associated natural frequency. 

Barring small neighborhoods of (semi)nodal points, the frequency response anywhere in the cylinder should be qualitatively the same and this is indeed borne out by numerical experiments. Given any representative point, each standing-wave excitation generates a series of resonances that are in correspondence with a subset of the natural frequencies of the simply-supported cylinder. Quantitative differences in the detailed features of the resonances (for example, their shapes and widths) are of course observed as the representative point is varied, but these differences are not usually of interest in steady-state vibration analyses of lossless isotropic elastic solids. 

\section{Supplementary Examples}\label{NumEG2}

As supplementary examples, we study the frequency response of three different cylinders, each possessing the same geometry and Poisson ratio $\nu=0.300$ but differing in their mass densities and Young moduli. The cylinder geometry is fixed as follows: $L=0.500$~m, $R_1=0.050$~m, and $R_2=0.150$~m. Thus, the mean radius $R\equiv(R_1+R_2)/2=0.100$~m, the thickness-to-radius ratio $h/R=1.00$, and the length-to-radius ratio $L/R=5.00$. The mass densities and Young moduli of the three cylinders are given in Table \ref{matprops3}. 

\begin{table}[h]
\centering
\begin{tabular}{ccc}
    \hline  \hline
     ~~Cylinder Material~~ & ~~$\rho$ ($\text{kg/m}^3$)~~ &  ~~$E$ (GPa)~~ \\ \hline
     Cadmium & 8650 & 50 \\
     Ruthenium &12370 & 447 \\
     Rhenium & 21020 & 463 \\ \hline \hline 
\end{tabular}
\caption{Three different cylinder materials having the same Poisson ratio $\nu=0.300$.}
\label{matprops3}
\end{table}

\begin{figure}[h]
\vspace*{0.5cm}
\centering 
\scalebox{0.6}{\includegraphics*{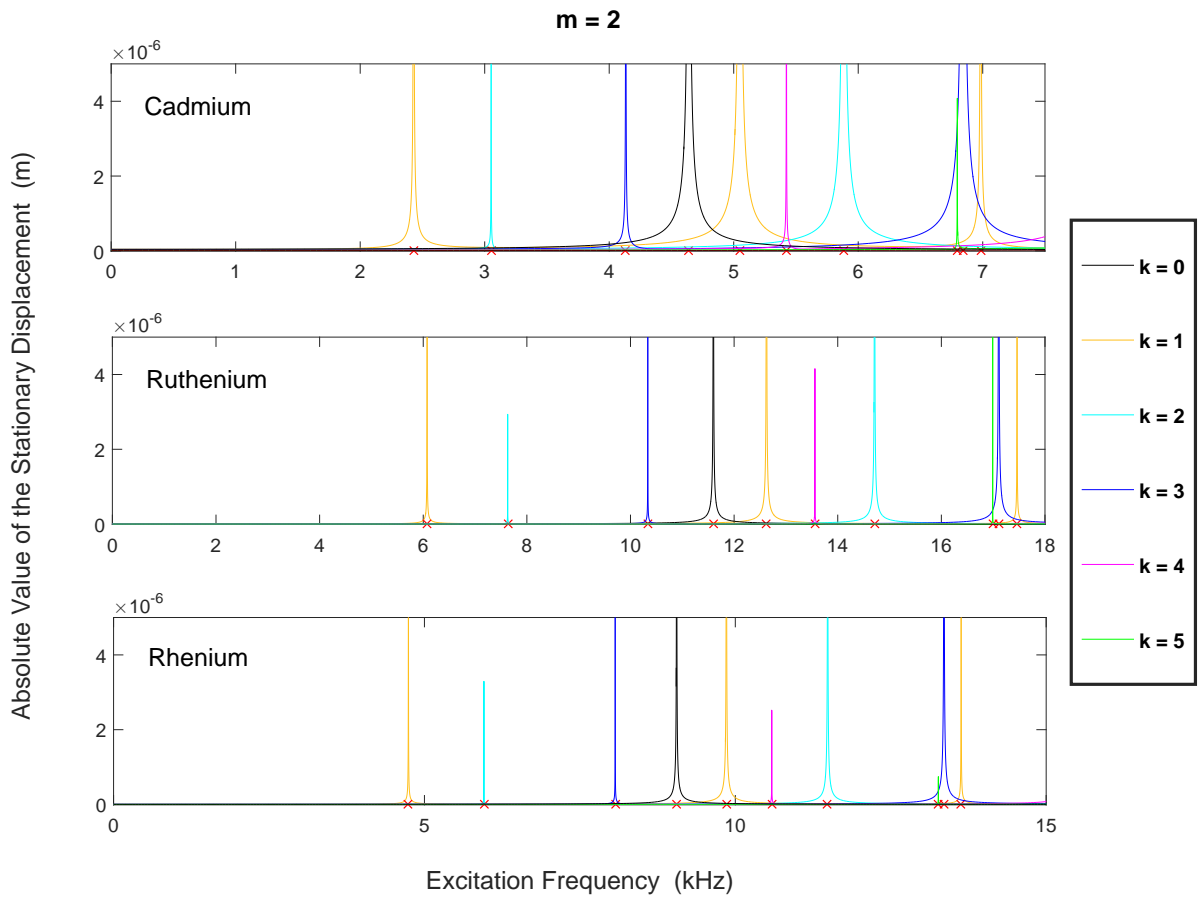}}
%\vspace*{-1.25cm}
\caption{\label{MLmeq2} Responses at the point $(r,\theta,z)=((R_1+R_2)/2,\pi/5,L/7)$ to various standing-wave excitations with wave numbers $m$ and $k$ as indicated for three different hollow cylinders. The cylinders possess identical geometries and Poisson ratios (see text) but differ in their mass densities and Young moduli (c.f., Table \ref{matprops3}). Natural frequencies pertinent to each case are marked by `$\mathsf{X}$'s on the frequency axis of each subplot.}
\end{figure}

For each of the three above-defined cylinders, the displacement responses at the point $(r,\theta,z)=((R_1+R_2)/2,\pi/5,L/7)$ to six different standing-wave excitations are shown in Fig.~\ref{MLmeq2}. Note that the displacement responses to the different standing-wave excitations are overlaid on each subplot with the understanding that they correspond to separate excitation cases. So, it should be understood that, for instance, the orange curve is the response to an excitation with wave numbers $(m,k)=(2,1)$, whereas the dark blue curve is the response to an excitation with wave numbers $(m,k)=(2,3)$. Natural frequencies pertinent to each case were computed using free-vibration frequency data obtained from Ref.~\cite{Arm69}, and as before, these frequencies are marked by `$\mathsf{X}$'s on the frequency axis of each subplot. It is interesting to note that each excitation excites the \emph{same} resonant modes independent of both the mass density and stiffness of the cylinder. For example, the fourth mode (corresponding to natural frequency $f^{(2)}_4$) is always excited by a standing-wave excitation with longitudinal wave number $k=0$, whereas the fifth mode (corresponding to natural frequency $f^{(2)}_5$) is always excited by a standing-wave excitation with longitudinal wave number $k=1$. Although not shown here, the same conclusion is reached when considering other non-zero circumferential wave numbers (i.e., $m\neq2$). 

\section{Numerical Epilogue: A Comment on Mode Orthogonality}

It is worthwhile to comment in more detail on the subtle and perhaps counterintuitive nature of the result that each harmonic standing-wave excitation excites many resonant modes. The excitations in the present problem are pure 2D boundary stresses that vary harmonically in the circumferential and axial directions. As such, these boundary stresses are characterized by two parameters: the circumferential wave number $m$ and the axial wave number $k$. The resonant modes of the cylinder, on the other hand, are three-dimensional, and hence cannot be uniquely specified using only the wave numbers $m$ and $k$. The key to understanding the preceding numerical results is to recognize that the wave numbers $m$ and $k$ are insufficient for indexing (i.e., uniquely classifying) all the different vibration modes of a thick-walled cylinder (a fundamental problem that has been previously discussed in the context of free vibrations in Refs.~\cite{Wills96,Wills98,Wills02}). For any given fixed values of $m$ and $k$ (and assuming that all other parameters are also fixed), there exists a countably-infinite set $\mathcal{M}(m,k)=\left\{\mathcal{M}_i(m,k): i\in\mathbb{Z}^+\right\}$ of physically distinct resonant modes each possessing a unique shape composed of $m$ full (cosine) waves around the circumference of the cylinder and $k$ half (sine) waves along the axis of the cylinder.\footnote{This is true by virtue of the fact that, for a simply-supported cylinder, there exists a countably-infinite set of unique free-vibration modes for any given fixed values of $m$ and $k$ \cite{Arm69}.} A resonant mode $\mathcal{M}_i(m,k)\in\mathcal{M}(m,k)$ is excited when the excitation frequency $f$ is at (or close to) the mode's respective resonant frequency, which, say, is equal to the natural frequency $f^{(m)}_j$. It is important to note that there is no formal correspondence between the integers $i$, $j$, and $k$, and for no reason should they be expected to possess equal values. If the values of $m$ and $k$ are fixed such that $(m,k)=(m^*,k^*)$ in excitations (\ref{strssrr2})-(\ref{strssrz2}), then these excitations will necessarily excite all of the modes in $\mathcal{M}(m^*,k^*)$ (each at its respective resonant frequency) because all of the modes in $\mathcal{M}(m^*,k^*)$ have wave numbers $m=m^*$ and $k=k^*$ (by definition). Despite the fact that all members of $\mathcal{M}(m^*,k^*)$ are characterized by the same circumferential and axial wave numbers, the constituent modes are: (i) excited at different frequencies; (ii) physically distinct (i.e., possess unique shapes); and most importantly (iii) linearly decoupled (i.e., no constituent mode is a linear combination of other constituent modes). In short, all members of any $\mathcal{M}(m,k)$ are orthogonal despite their common wave numbers $m$ and $k$. Hence, the observed numerical results do not violate mode orthogonality.  

\begin{figure}[h]
\vspace*{0.5cm}
\centering 
\scalebox{0.37}{\includegraphics*{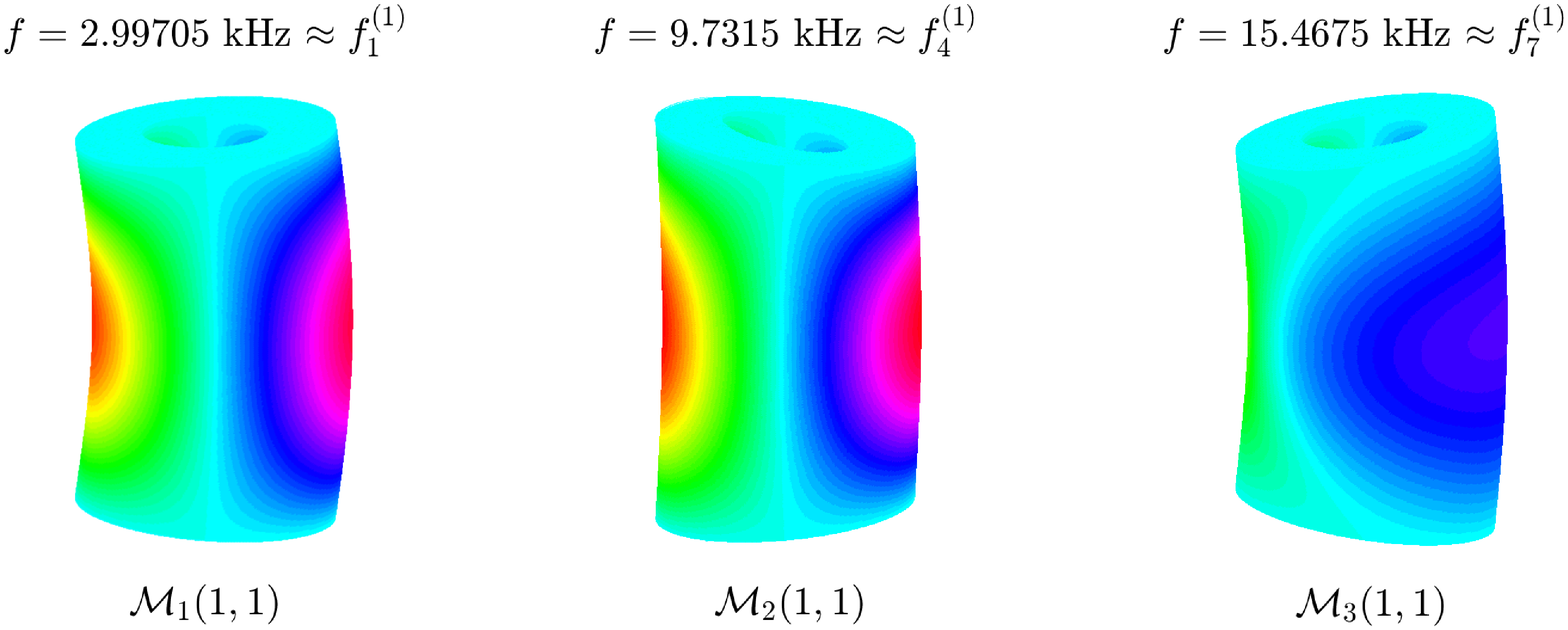}} \par 
\vspace*{0.4cm}
\scalebox{0.42}{\includegraphics*{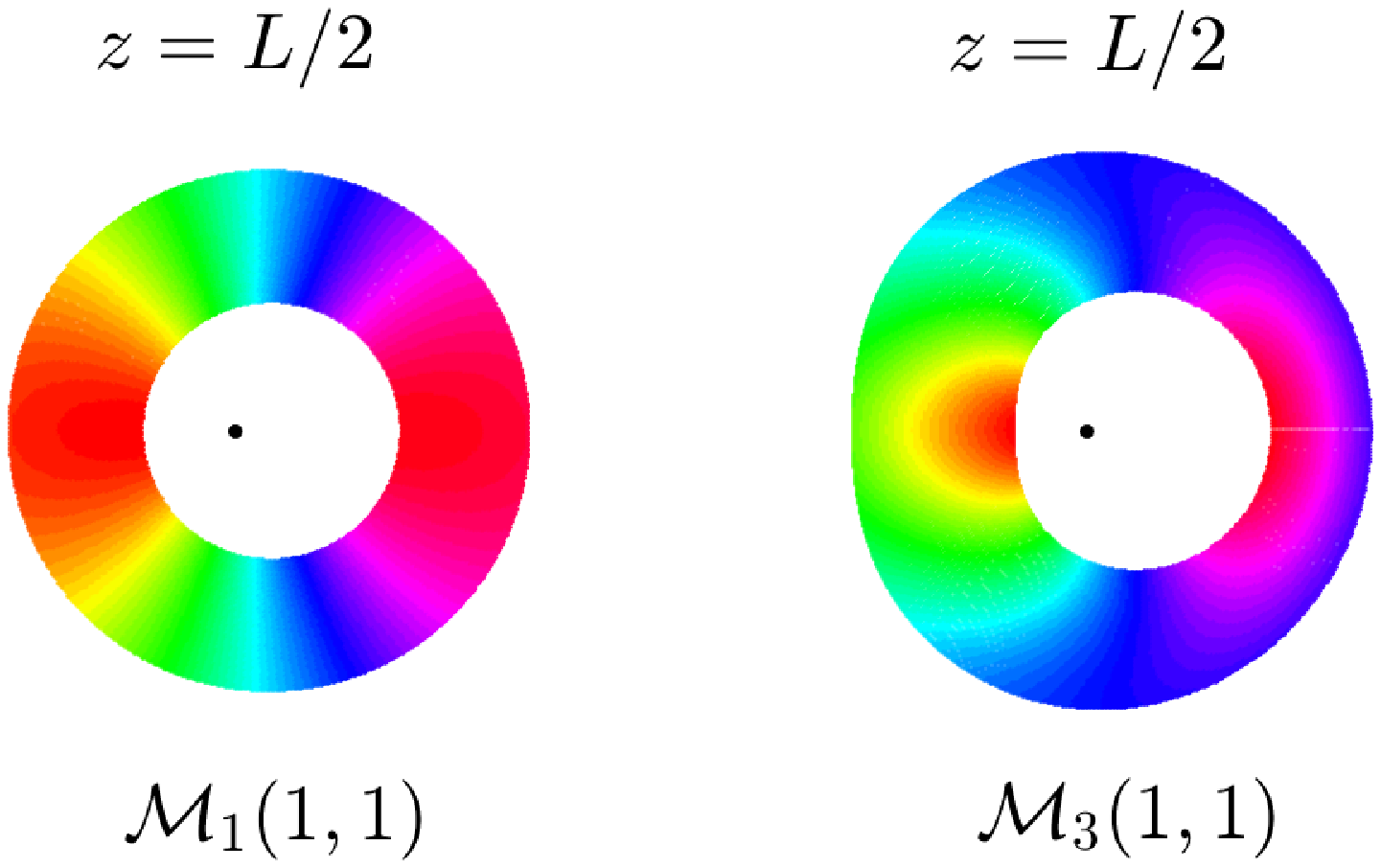}}
\vspace*{-0.1cm}
\caption{\label{MSeg} (Top Row) Stationary shapes of the first three members of $\mathcal{M}(1,1)$ for the cylinder considered in Example 1. The respective excitation frequencies are as indicated. (Bottom Row) Aerial views of the shapes of $\mathcal{M}_1(1,1)$ and $\mathcal{M}_3(1,1)$ at $z=L/2$. For reference, the center point of the undeformed free cylinder is (in each case) marked by a black dot.}
\end{figure}

To give a numerical example, we revisit Example 1 and explicitly compute the (stationary) shapes of the first three resonant modes in $\mathcal{M}(1,1)$. For these computations, all parameters are set to the values given in Section \ref{numerica} except the stress amplitudes $\{\mathcal{A},\mathcal{B},\mathcal{C},\mathcal{D},\mathcal{E},\mathcal{F}\}$, which (for simplicity) are here set as follows: $\mathcal{A}=\mathcal{C}=\mathcal{D}=\mathcal{F}=0$ and $\mathcal{B}=\mathcal{E}=0.500$ MPa. The results, obtained from use of our exact $k\neq0$ solution (Section \ref{solntoBVP}), are shown in Fig.~\ref{MSeg}. As validation of these results, we note that the shapes of $\mathcal{M}_1(1,1)$ and $\mathcal{M}_2(1,1)$ are fully consistent with the corresponding free-vibration mode shapes given in Ref.~\cite{Ye14}.\footnote{Specifically, the first and fourth modes (respectively) shown in Fig.~4 of Ref.~\cite{Ye14} in the ``S-S n=1'' case. Note that the circumferential wave number is denoted by the letter $n$ in Ref.~\cite{Ye14}.} To further expose the differences in the shapes of $\mathcal{M}_1(1,1)$ and $\mathcal{M}_3(1,1)$, we display their respective shapes in the mid-plane $z=L/2$ in the lower panel of Fig.~\ref{MSeg}. Clearly, the three modes shown in Fig.~\ref{MSeg} are orthogonal, and as expected, the mode numbers $i$ and $j$ (as defined above) generally differ from the axial wave number $k$ (see Table \ref{TabLmodenums}). 

\begin{table}[h]
\centering
\begin{tabular}{cccc}
    \hline \hline 
    Mode & ~~~$i$~~~ & ~~~$j$~~~ & ~~~$k$~~~ \\ \hline
    $\mathcal{M}_1(1,1)$ & $1$ & $1$ & $1$ \\ 
    $\mathcal{M}_2(1,1)$ & $2$ & $4$ & $1$ \\
    $\mathcal{M}_3(1,1)$ & $3$ & $7$ & $1$ \\  
    \hline \hline
\end{tabular}
\caption{Values of mode numbers $i$ and $j$ (as defined in the text) and axial wave number $k$ for the three modes shown in Fig.~\ref{MSeg}.}
\label{TabLmodenums}
\end{table}

Given the above comments, it is clear that no resonant mode can be individually excited by excitations (\ref{strssrr2})-(\ref{strssrz2}) for any given fixed values of the wave numbers $m$ and $k$. 
%This actually furnishes a specific example of a more general (but not widely recognized) phenomenon, namely, that steady-state \emph{non-volumetric} excitations of a thick-walled cylinder (consisting of pure boundary stresses, pure boundary displacements, or combinations thereof) cannot excite individual resonant modes. 
 
\section{Conclusion}\label{conclus}

In summary, we have studied analytically the linear elastodynamic response of a simply-supported isotropic thick-walled hollow elastic circular cylinder subjected to 2D harmonic standing-wave excitations on its curved surfaces. Our mathematical formulation employed the exact 3D theory of linear elasticity. An exact semi-analytical solution for the steady-state displacement field of the cylinder was constructed using recently-obtained parametric solutions to the Navier-Lam\'{e} equation. In order to provide a solution that involves only real-valued Bessel functions, the problem was solved separately in three distinct parameter regimes involving the excitation frequency. In each case, application of the standing-wave boundary conditions generates a parameter-dependent $6\times6$ linear system, which can be solved numerically in order to study the parametric responses of the cylinder's displacement field under various conditions. Standard Bessel function identities were also used to cast the radial part of the solution (which involves Bessel functions of the first and second kinds) in a derivative-free form thereby making the obtained semi-analytical solution apt for numerical computation. The method of solution, which exploits known solutions to the Navier-Lam\'{e} equation \cite{usB1}, demonstrates a general approach that can be applied to solve other elastodynamic forced-response problems involving isotropic, open or closed, solid or hollow, elastic cylinders with simply-supported or other end  conditions. 

As an application, and considering several numerical examples, the obtained solution was used to determine the steady-state frequency response in some low-order excitation cases. In each case, the solution generates a series of resonances that are in perfect correspondence with a unique subset of the natural frequencies of the \emph{simply-supported} cylinder. It is worth emphasizing that \emph{each} standing-wave excitation generates a unique \emph{series} of resonances (of varying widths), as opposed to generating just a single resonance. Put another way, each standing-wave excitation excites a set of (orthogonal) resonant modes as opposed to exciting only one resonant mode. Unfortunately, for any given value of the circumferential wave number $m$ (and assuming the values of all cylinder parameters are fixed), there is no way of predicting which modes will be excited given the value of the longitudinal wave number $k$. 

While the above numerical results provide important physical insight, they are cursory in the sense that many fundamental aspects of the studied problem remain numerically unexplored. For instance, there is the fundamental problem of the ``resonance dynamics'', that is, how the resonances generated by any given standing-wave excitation evolve in frequency space under the variation of a material or geometric parameter. The resonant mode shapes and their evolution under parametric variation(s) is another fundamental aspect requiring systematic investigation, and (as mentioned in the Introduction) the ultimate goal is to tackle the more general forced-vibration problem involving arbitrary \emph{asymmetric} surface excitations -- a problem for which the present work provides the necessary foundation. We hope to explore these topics in future publications. 

\appendix 

\section{General Axisymmetric Solution Using the ERP Field Equations}\label{appendx2}

In this appendix, we show how our general axisymmetric solution may be reproduced from the general ERP field equations. For brevity, we shall here only provide the derivation for the Case 1 sub-solution. The sub-solutions in the other two cases may be similarly reproduced using the same logic. 

\subsection{General ERP Field Equations}

According to ERP \cite{Ebenezer15}, the following is an exact \emph{axisymmetric} solution to Eq.~(\ref{NLE}), for arbitrary values of $k_{rn}~(n=1,2,3,\ldots,N_r)$ and $k_{zn}~(n=1,2,3,\ldots,N_z)$: 
\begin{subequations}\label{Ebensoln}
\begin{eqnarray}\label{EbensolnT}
\left[\begin{array}{c} u^{(ERP)}_z \\ u^{(ERP)}_r \end{array} \right] = \left[\begin{array}{c}
u^{(1)}_z \\ u^{(1)}_r \end{array} \right] + \left[\begin{array}{c}
u^{(2)}_z \\ u^{(2)}_r \end{array} \right] + \left[\begin{array}{c}
u^{(3)}_z \\ u^{(3)}_r \end{array} \right], 
\end{eqnarray}
where
\begin{eqnarray}\label{EbensolnP1}
\left[\begin{array}{c} u^{(1)}_z \\ u^{(1)}_r \end{array} \right] = \left[\begin{array}{c} P\cos(\Omega_1z)+{\displaystyle\sum_{n=1}^{N_r}}{\displaystyle\sum_{s=1}^2}P_{ns}C_0(k_{rn}r)\cos(k_{zns}z) \\ {\displaystyle\sum_{n=1}^{N_r}}{\displaystyle\sum_{s=1}^2}P_{ns}\psi_{ns}C_1(k_{rn}r)\sin(k_{zns}z) \end{array} \right],  
\end{eqnarray}
\begin{eqnarray}\label{EbensolnP2}
\left[\begin{array}{c} u^{(2)}_z \\ u^{(2)}_r \end{array} \right] = \left[\begin{array}{c} QJ_0(\Omega_2r)+{\displaystyle\sum_{n=1}^{N_z}}{\displaystyle\sum_{s=1}^2}Q_{ns}J_0(k_{rns}r)\cos(k_{zn}z) \\ {\displaystyle\sum_{n=1}^{N_z}}{\displaystyle\sum_{s=1}^2}Q_{ns}\chi_{ns}J_1(k_{rns}r)\sin(k_{zn}z) \end{array} \right],  
\end{eqnarray}
\begin{eqnarray}\label{EbensolnP3}
\left[\begin{array}{c} u^{(3)}_z \\ u^{(3)}_r \end{array} \right] = \left[\begin{array}{c} RY_0(\Omega_2r)+{\displaystyle\sum_{n=1}^{N_z}}{\displaystyle\sum_{s=1}^2}R_{ns}Y_0(k_{rns}r)\cos(k_{zn}z) \\ {\displaystyle\sum_{n=1}^{N_z}}{\displaystyle\sum_{s=1}^2}R_{ns}\chi_{ns}Y_1(k_{rns}r)\sin(k_{zn}z) \end{array} \right],  
\end{eqnarray}
\begin{equation}
\Omega_1=\sqrt{\rho\omega^2\over(\lambda+2\mu)}, \quad \Omega_2=\sqrt{\rho\omega^2\over\mu},
\end{equation}
\begin{equation}\label{LCsJsYs}
C_0(k_{rn}r)=J_0(k_{rn}r)+\zeta_nY_0(k_{rn}r), \quad C_1(k_{rn}r)=J_1(k_{rn}r)+\zeta_nY_1(k_{rn}r), 
\end{equation}
$P$, $Q$, $R$, $\{P_{ns}:n=1,2,3,\ldots,N_r;~s=1,2\}$, $\{Q_{ns}:n=1,2,3,\ldots,N_z;~s=1,2\}$, $\{R_{ns}:n=1,2,3,\ldots,N_z;~s=1,2\}$ are arbitrary constants (determined by the specific excitation), and the remaining (frequency-dependent) constants are given by: 
\begin{equation}
k_{zn1}=\sqrt{{\rho\omega^2\over(\lambda+2\mu)}-k^2_{rn}}, \quad k_{zn2}=\sqrt{{\rho\omega^2\over\mu}-k^2_{rn}}, \quad n=1,2,3,\ldots,N_r
\end{equation}
\begin{equation}
k_{rn1}=\sqrt{{\rho\omega^2\over(\lambda+2\mu)}-k^2_{zn}}, \quad k_{rn2}=\sqrt{{\rho\omega^2\over\mu}-k^2_{zn}}, \quad n=1,2,3,\ldots,N_z
\end{equation}
\begin{equation}
\psi_{n1}={k_{rn}\over k_{zn1}}, \quad \psi_{n2}=-{k_{zn2}\over k_{rn}}, \quad n=1,2,3,\ldots,N_r
\end{equation}
\begin{equation}
\chi_{n1}=-{k_{rn1}\over k_{zn}}, \quad \chi_{n2}={k_{zn}\over k_{rn2}}, \quad n=1,2,3,\ldots,N_z.
\end{equation}
\end{subequations}
The non-arbitrary constants $\{\zeta_n:n=1,2,3,\ldots,N_r\}$ in (\ref{LCsJsYs}) are chosen so as to satisfy certain conditions (see Ref.~\cite{Ebenezer15} for details). Since these constants will have no relevance in the analyses that follow, it is sufficient for our purposes to leave them unspecified. 

The components of the stress field corresponding to the displacement field (\ref{Ebensoln}) are then as follows \cite{Ebenezer15}: 
\begin{eqnarray}\label{DDEsigmaRR}
\sigma^{(ERP)}_{rr}&=&-P\Omega_1\lambda\sin(\Omega_1z) \nonumber \\
&+&{\displaystyle\sum_{n=1}^{N_r}}{\displaystyle\sum_{s=1}^2}P_{ns}\left\{\Big[(\lambda+2\mu)\psi_{ns}k_{rn}-\lambda k_{zns}\Big]C_0(k_{rn}r)-{2\mu\over r}\psi_{ns}C_1(k_{rn}r)\right\}\sin(k_{zns}z) \nonumber \\
&+&{\displaystyle\sum_{n=1}^{N_z}}{\displaystyle\sum_{s=1}^2}Q_{ns}\left\{\Big[(\lambda+2\mu)\chi_{ns}k_{rns}-\lambda k_{zn}\Big]J_0(k_{rns}r)-{2\mu\over r}\chi_{ns}J_1(k_{rns}r)\right\}\sin(k_{zn}z) \nonumber \\ 
&+&{\displaystyle\sum_{n=1}^{N_z}}{\displaystyle\sum_{s=1}^2}R_{ns}\left\{\Big[(\lambda+2\mu)\chi_{ns}k_{rns}-\lambda k_{zn}\Big]Y_0(k_{rns}r)-{2\mu\over r}\chi_{ns}Y_1(k_{rns}r)\right\}\sin(k_{zn}z), \nonumber \\ 
\end{eqnarray}
\begin{eqnarray}\label{DDEsigmaZZ}
\sigma^{(ERP)}_{zz}&=&-P\Omega_1(\lambda+2\mu)\sin(\Omega_1z) \nonumber \\
&+&{\displaystyle\sum_{n=1}^{N_r}}{\displaystyle\sum_{s=1}^2}P_{ns}\Big[-(\lambda+2\mu)k_{zns}+\lambda\psi_{ns}k_{rn}\Big]C_0(k_{rn}r)\sin(k_{zns}z) \nonumber \\
&+&{\displaystyle\sum_{n=1}^{N_z}}{\displaystyle\sum_{s=1}^2}Q_{ns}\Big[-(\lambda+2\mu)k_{zn}+\lambda\chi_{ns}k_{rns}\Big]J_0(k_{rns}r)\sin(k_{zn}z) \nonumber \\
&+&{\displaystyle\sum_{n=1}^{N_z}}{\displaystyle\sum_{s=1}^2}R_{ns}\Big[-(\lambda+2\mu)k_{zn}+\lambda\chi_{ns}k_{rns}\Big]Y_0(k_{rns}r)\sin(k_{zn}z),
\end{eqnarray}
and 
\begin{eqnarray}\label{DDEsigmaRZ}
\sigma^{(ERP)}_{rz}&=&-Q\Omega_2\mu J_1(\Omega_2r)-R\Omega_2\mu Y_1(\Omega_2r) \nonumber \\
&+&\mu{\displaystyle\sum_{n=1}^{N_r}}{\displaystyle\sum_{s=1}^2}P_{ns}\Big[-k_{rn}+\psi_{ns}k_{zns}\Big]C_1(k_{rn}r)\cos(k_{zns}z) \nonumber \\
&+&\mu{\displaystyle\sum_{n=1}^{N_z}}{\displaystyle\sum_{s=1}^2}Q_{ns}\Big[-k_{rns}+\chi_{ns}k_{zn}\Big]J_1(k_{rns}r)\cos(k_{zn}z) \nonumber \\
&+&\mu{\displaystyle\sum_{n=1}^{N_z}}{\displaystyle\sum_{s=1}^2}R_{ns}\Big[-k_{rns}+\chi_{ns}k_{zn}\Big]Y_1(k_{rns}r)\cos(k_{zn}z).
\end{eqnarray}
As noted in ERP \cite{Ebenezer15}, the harmonic time dependence has been dropped (for convenience) from all field equations. 

\subsection{Reduction and Equivalency of the ERP Field Equations}\label{ReducEquivERP}

To enforce the vanishing of the longitudinal stress component (\ref{DDEsigmaZZ}) at the ends of the cylinder (i.e., to satisfy condition (\ref{SSBCs3})), we set the arbitrary ERP constant $k_{zn}=(n\pi/L)$ and prescribe $P_{ns}=0,~\forall n,s$ in Eq.~(\ref{DDEsigmaZZ}) and thereby in all of the remaining ERP field equations. The values of the constants $\psi_{ns}$, $k_{rn}$, and $k_{zns}$ appearing in the associated summands are subsequently irrelevant. Unless the parameters are such that ${\rho\omega^2/(\lambda+2\mu)}=\left({n\pi/L}\right)^2$ (a singular case that we have ignored, see comments at the end of Section \ref{generalKneq0}), we can also prescribe $P=0$ in (\ref{DDEsigmaZZ}) and thereby in (\ref{DDEsigmaRR}) and (\ref{EbensolnP1}). Hence, under the condition that ${\rho\omega^2/(\lambda+2\mu)}\neq\left({n\pi/L}\right)^2$, the sub-solution (\ref{EbensolnP1}) is zero, i.e., $\left(u^{(1)}_z,u^{(1)}_r\right)=(0,0)$. The general displacement field that we seek is thus entirely contained in sub-solutions (\ref{EbensolnP2}) and (\ref{EbensolnP3}). 
 
Anticipating that we shall only require the individual harmonics of the stresses (\ref{DDEsigmaRR})-(\ref{DDEsigmaRZ}) and thereby only the corresponding individual harmonics of the displacement field, each sum over $n$ in Eqs.~(\ref{Ebensoln})-(\ref{DDEsigmaRZ}) collapses to a single term that depends on the prescribed value of $n$. Universally replacing the index $n$ (in the ERP field equations) by $k$ (the longitudinal wave number in our problem) then yields the following for the frequency-dependent ERP constants:
\begin{eqnarray}\label{DDEconst1}
k_{rn	1} \longrightarrow k_{rk1}=\sqrt{{\rho\omega^2\over(\lambda+2\mu)}-\left({k\pi\over L}\right)^2}=\left\{\begin{array}{rrr}
             i\alpha_1 & ~~\text{Case~1} \\
             \alpha_1 & ~~\text{Case~2} \\
             i\alpha_1 & ~~\text{Case~3} \\
             \end{array} \right., 
\end{eqnarray}
\begin{eqnarray}\label{DDEconst2}
k_{rn	2} \longrightarrow k_{rk2}=\sqrt{{\rho\omega^2\over\mu}-\left({k\pi\over L}\right)^2}=\left\{\begin{array}{rrr}
             i\alpha_2 & ~~\text{Case~1} \\
             \alpha_2 & ~~\text{Case~2} \\
             \alpha_2 & ~~\text{Case~3} \\
             \end{array} \right., 
\end{eqnarray}
\begin{eqnarray}\label{DDEconst3}
\chi_{n1} \longrightarrow \chi_{k1}=-{k_{rk1}\over k_{zk}}=\left\{\begin{array}{rrr}
             \alpha_1/i\left({k\pi\over L}\right) & ~~\text{Case~1} \\
             -\alpha_1/\left({k\pi\over L}\right) & ~~\text{Case~2} \\
             \alpha_1/i\left({k\pi\over L}\right) & ~~\text{Case~3} \\
             \end{array} \right., 
\end{eqnarray}
\begin{eqnarray}\label{DDEconst4}
\chi_{n2} \longrightarrow \chi_{k2}={k_{zk}\over k_{rk2}}=\left\{\begin{array}{rrr}
             \left({k\pi\over L}\right)/i\alpha_2 & ~~\text{Case~1} \\
             \left({k\pi\over L}\right)/\alpha_2 & ~~\text{Case~2} \\
             \left({k\pi\over L}\right)/\alpha_2 & ~~\text{Case~3} \\
             \end{array} \right., 
\end{eqnarray}
where the final equalities in (\ref{DDEconst1})-(\ref{DDEconst4}) are with reference to constants $\{\alpha_1,\alpha_2\}$ as defined by Eqs.~(\ref{alphasEG}), (\ref{alphasEG5}), and (\ref{alphasEG3}), for Cases 1, 2, and 3, respectively. 

\subsubsection{Displacement Components}

For Case 1, the radial component of the ERP displacement field, which is the sum of the non-vanishing radial components $u^{(2)}_r$ and $u^{(3)}_r$ in sub-solutions (\ref{EbensolnP2}) and (\ref{EbensolnP3}), reduces as follows:
\begin{eqnarray}\label{DDERAD2}
u^{(ERP)}_r&=&\left\{{\displaystyle\sum_{s=1}^2}\chi_{ks}\Big[Q_{ks}J_1(k_{rks}r)+R_{ks}Y_1(k_{rks}r)\Big]\right\}\sin(k_{zk}z) \nonumber \\ 
&=&\left\{{\alpha_1\over\left({k\pi\over L}\right)}\Big[Q_{k1}i^{-1}J_1(i\alpha_1r)+R_{k1}i^{-1}Y_1(i\alpha_1r)\Big]\right. \nonumber \\
&&~\left.+~{\left({k\pi\over L}\right)\over\alpha_2}\Big[Q_{k2}i^{-1}J_1(i\alpha_2r)+R_{k2}i^{-1}Y_1(i\alpha_2r)\Big]\right\}\sin\left({k\pi\over L}z\right) \nonumber \\
&=&\left\{{\alpha_1\over\left({k\pi\over L}\right)}\Big[(Q_{k1}+iR_{k1})i^{-1}J_1(i\alpha_1r)+{2\over\pi}R_{k1}K_1(\alpha_1r)\Big]\right. \nonumber \\
&&~\left.+~{\left({k\pi\over L}\right)\over\alpha_2}\Big[(Q_{k2}+iR_{k2})i^{-1}J_1(i\alpha_2r)+{2\over\pi}R_{k2}K_1(\alpha_2r)\Big]\right\}\sin\left({k\pi\over L}z\right) \nonumber \\ 
&=&\left\{{\alpha_1\over\left({k\pi\over L}\right)}\Big[\widetilde{Q}_{k1}I_1(\alpha_1r)+\widetilde{R}_{k1}K_1(\alpha_1r)\Big]\right. \nonumber \\
&&~\left.+~{\left({k\pi\over L}\right)\over\alpha_2}\Big[\widetilde{Q}_{k2}I_1(\alpha_2r)+\widetilde{R}_{k2}K_1(\alpha_2r)\Big]\right\}\sin\left({k\pi\over L}z\right) \nonumber \\ 
&=&\left\{{\alpha_1}\Big[A_1I_1(\alpha_1r)-B_1K_1(\alpha_1r)\Big]+{\alpha_2}\Big[A_2I_1(\alpha_2r)-B_2K_1(\alpha_2r)\Big]\right\}\sin\left({k\pi\over L}z\right),~~~~~
\end{eqnarray}
where relations (\ref{DDEconst1})-(\ref{DDEconst4}) have been employed in obtaining the second equality in (\ref{DDERAD2}), the definition of the modified Bessel function of the second kind
\begin{equation}\label{defnKBessel}
K_n(x)={\pi\over2}i^{n+1}\big[J_n(ix)+iY_n(ix)\big], \quad n\in\mathbb{N}
\end{equation}
in obtaining the third equality, and the fundamental definition of the modified Bessel function of the first kind   
\begin{equation}\label{BFiden}
I_p(x)\equiv i^{-p}J_p(ix), \quad p\in\mathbb{R} 
\end{equation}
in obtaining the fourth equality. The last equality in (\ref{DDERAD2}) ensues upon replacement of the arbitrary constants 
\begin{equation}
\widetilde{Q}_{ks}\equiv Q_{ks}+iR_{ks}, \quad \widetilde{R}_{ks}\equiv(2/\pi)R_{ks} \quad \quad (s=1,2)
\end{equation}
by a new set of arbitrary constants $\left\{A_1,A_2,B_1,B_2\right\}$ as follows:
\begin{subequations}\label{correspconstsC1}
\begin{eqnarray}\label{correspconstsC1P1}
\widetilde{Q}_{ks} \longrightarrow \left({k\pi\over L}\right)A_s\gamma_s=\left\{\begin{array}{lll}
             \left({k\pi\over L}\right)A_1 & ~~\text{if}~s=1 \\
             {\alpha^2_2A_2/\left({k\pi\over L}\right)} & ~~\text{if}~s=2 \\
             \end{array} \right., 
\end{eqnarray} 
\begin{eqnarray}\label{correspconstsC1P2}
\widetilde{R}_{ks} \longrightarrow -\left({k\pi\over L}\right)B_s\gamma_s=\left\{\begin{array}{lll}
             -\left({k\pi\over L}\right)B_1 & ~~\text{if}~s=1 \\
             {-\alpha^2_2B_2/\left({k\pi\over L}\right)} & ~~\text{if}~s=2 \\
             \end{array} \right.. 
\end{eqnarray} 
\end{subequations} 

Using the same logic (and again considering Case 1), the axial component of the ERP displacement field, which is the sum of the non-vanishing axial components $u^{(2)}_z$ and $u^{(3)}_z$ in sub-solutions (\ref{EbensolnP2}) and (\ref{EbensolnP3}), similarly reduces as follows:
\begin{eqnarray}\label{DDEAXL2}
u^{(ERP)}_z&=&QJ_0(\Omega_2r)+RY_0(\Omega_2r)+\left\{{\displaystyle\sum_{s=1}^2}\Big[Q_{ks}J_0(k_{rks}r)+R_{ks}Y_0(k_{rks}r)\Big]\right\}\cos(k_{zk}z) \nonumber \\ 
&=&QJ_0(\Omega_2r)+\Big[Q_{k1}J_0(i\alpha_1r)+Q_{k2}J_0(i\alpha_2r)\Big]\cos\left({k\pi\over L}z\right) \nonumber \\ 
&&~+~RY_0(\Omega_2r)+\Big[R_{k1}Y_0(i\alpha_1r)+R_{k2}Y_0(i\alpha_2r)\Big]\cos\left({k\pi\over L}z\right) \nonumber \\
&=&QJ_0(\Omega_2r)+\Big[(Q_{k1}+iR_{k1})J_0(i\alpha_1r)+(Q_{k2}+iR_{k2})J_0(i\alpha_2r)\Big]\cos\left({k\pi\over L}z\right) \nonumber \\ 
&&~+~RY_0(\Omega_2r)-{2\over\pi}\Big[R_{k1}K_0(\alpha_1r)+R_{k2}K_0(\alpha_2r)\Big]\cos\left({k\pi\over L}z\right) \nonumber \\
&=&QJ_0(\Omega_2r)+\Big[\widetilde{Q}_{k1}I_0(\alpha_1r)+\widetilde{Q}_{k2}I_0(\alpha_2r)\Big]\cos\left({k\pi\over L}z\right) \nonumber \\
&&~+~RY_0(\Omega_2r)-\Big[\widetilde{R}_{k1}K_0(\alpha_1r)+\widetilde{R}_{k2}K_0(\alpha_2r)\Big]\cos\left({k\pi\over L}z\right) \nonumber \\
&=&\Big[QJ_0(\Omega_2r)+RY_0(\Omega_2r)\Big]+\left({k\pi\over L}\right)\bigg\{{\gamma_1}\Big[A_1I_0(\alpha_1r)+B_1K_0(\alpha_1r)\Big] \nonumber \\
&&~+~{\gamma_2}\Big[A_2I_0(\alpha_2r)+B_2K_0(\alpha_2r)\Big]\bigg\}\cos\left({k\pi\over L}z\right). 
\end{eqnarray}
The first bracketed term in the final line of (\ref{DDEAXL2}) is equivalent to the axial component of (\ref{gensolnkeq0}) in the special $m=0$ case. Comparing (\ref{DDERAD2}) with (\ref{SolBVP2C1meq01}) and the cosine term of (\ref{DDEAXL2}) with (\ref{SolBVP2C1meq02}), we see that, in Case 1, the displacement components obtained from the ERP method are identical to the displacement components obtained from our method of solution in the special $m=0$ case. Using relations (\ref{DDEconst1})-(\ref{DDEconst4}), definitions (\ref{defnKBessel})-(\ref{BFiden}), and analogs of substitutions (\ref{correspconstsC1}), and applying the same logic, it is straightforward to show that the equivalence holds for Cases 2 and 3 as well. Thus, in the special $m=0$ case, the general displacement field obtained from our method of solution is identical to the general axisymmetric displacement field obtained from the ERP method.  

\subsubsection{Stress Components}

For Case 1, the axisymmetric radial stress obtained from our method of solution is (omitting the $\sin(\omega t)$ factor): 
\begin{eqnarray}\label{OURsigmaRRC1meq0}
\sigma^{(SC)}_{rr}&=&\left\{{\displaystyle\sum_{s=1}^2}A_s\left[\beta_sI_0(\alpha_sr)-{2\mu\alpha_s\over r}I_1(\alpha_sr)\right]\right. \nonumber \\
&&~+~\left.{\displaystyle\sum_{s=1}^2}B_s\left[\beta_sK_0(\alpha_sr)+{2\mu\alpha_s\over r}K_1(\alpha_sr)\right]\right\} \sin\left({k\pi\over L}z\right), \quad
\end{eqnarray}
which is immediately obtained upon substituting $m=0$ into Eq.~(\ref{strsscmpRRC1}). 

Given the statements at the beginning of Section \ref{ReducEquivERP}, the ERP radial stress component (\ref{DDEsigmaRR}) reduces to the following:
\begin{eqnarray}\label{DDEsigmaRRreduce}
\sigma^{(ERP)}_{rr}&=&{\displaystyle\sum_{s=1}^2}Q_{ks}\left\{\Big[(\lambda+2\mu)\chi_{ks}k_{rks}-\lambda k_{zk}\Big]J_0(k_{rks}r)-{2\mu\over r}\chi_{ks}J_1(k_{rks}r)\right\}\sin(k_{zk}z) \nonumber \\
&+&{\displaystyle\sum_{s=1}^2}R_{ks}\left\{\Big[(\lambda+2\mu)\chi_{ks}k_{rks}-\lambda k_{zk}\Big]Y_0(k_{rks}r)-{2\mu\over r}\chi_{ks}Y_1(k_{rks}r)\right\}\sin(k_{zk}z). \nonumber \\ 
\end{eqnarray}
Employing relations (\ref{DDEconst1})-(\ref{DDEconst4}) and considering only Case 1, the frequency-dependent ERP constants in (\ref{DDEsigmaRRreduce}) simplify as follows:
\begin{subequations}\label{ERPconstsRADSTRSS}
\begin{equation}\label{ERPconstsRADSTRSSC1}
C_{k1}(\omega)\equiv(\lambda+2\mu)\chi_{k1}k_{rk1}-\lambda k_{zk}={(\lambda+2\mu)\alpha^2_1\over\left({k\pi\over L}\right)}-\lambda\left({k\pi\over L}\right)={\beta_1\over\left({k\pi\over L}\right)},
\end{equation}
\begin{equation}\label{ERPconstsRADSTRSSC2}
C_{k2}(\omega)\equiv(\lambda+2\mu)\chi_{k2}k_{rk2}-\lambda k_{zk}=2\mu\left({k\pi\over L}\right)={\beta_2\left({k\pi\over L}\right)\over\alpha^2_2},
\end{equation}
where the final equalities in (\ref{ERPconstsRADSTRSSC1})-(\ref{ERPconstsRADSTRSSC2}) are with reference to the (frequency-dependent) constants $\{\beta_1,\beta_2\}$ as defined by Eq.~(\ref{strsscmpRRC11B}). For future algebraic convenience, we also define the frequency-dependent constants $\{D_{k1}(\omega),D_{k2}(\omega)\}$ as follows:
\begin{equation}\label{ERPconstsRADSTRSSC3}
\chi_{k1}={\alpha_1\over\left({k\pi\over L}\right)}i^{-1}\equiv D_{k1}(\omega)i^{-1}, \quad \chi_{k2}={\left({k\pi\over L}\right)\over\alpha_2}i^{-1}\equiv D_{k2}(\omega)i^{-1},
\end{equation}
where the values of $\{\chi_{ks}: s=1,2\}$ in (\ref{ERPconstsRADSTRSSC3}) are again pertinent to Case 1. 
\end{subequations}
Using the compact notation of (\ref{ERPconstsRADSTRSS}) for the frequency-dependent ERP constants and keeping in mind that we are here only considering Case 1, the ERP radial stress component (\ref{DDEsigmaRRreduce}) then reduces as follows:
\begin{eqnarray}\label{DDEsigmaRRreduceF}
\sigma^{(ERP)}_{rr}&=&\left\{{\displaystyle\sum_{s=1}^2}Q_{ks}\left[C_{ks}(\omega)J_0(i\alpha_sr)-{2\mu\over r}D_{ks}(\omega)i^{-1}J_1(i\alpha_sr)\right]\right. \nonumber \\
&&~+~\left.{\displaystyle\sum_{s=1}^2}R_{ks}\left[C_{ks}(\omega)Y_0(i\alpha_sr)-{2\mu\over r}D_{ks}(\omega)i^{-1}Y_1(i\alpha_sr)\right]\right\} \sin\left({k\pi\over L}z\right) \nonumber \\ 
&=&\left\{{\displaystyle\sum_{s=1}^2}(Q_{ks}+iR_{ks})\left[C_{ks}(\omega)J_0(i\alpha_sr)-{2\mu\over r}D_{ks}(\omega)i^{-1}J_1(i\alpha_sr)\right]\right. \nonumber \\
&&~-~\left.{\displaystyle\sum_{s=1}^2}{2\over\pi}R_{ks}\left[C_{ks}(\omega)K_0(\alpha_sr)+{2\mu\over r}D_{ks}(\omega)K_1(\alpha_sr)\right]\right\} \sin\left({k\pi\over L}z\right) \nonumber \\ 
&=&\left\{{\displaystyle\sum_{s=1}^2}\widetilde{Q}_{ks}\left[C_{ks}(\omega)I_0(\alpha_sr)-{2\mu\over r}D_{ks}(\omega)I_1(\alpha_sr)\right]\right. \nonumber \\
&&~-~\left.{\displaystyle\sum_{s=1}^2}\widetilde{R}_{ks}\left[C_{ks}(\omega)K_0(\alpha_sr)+{2\mu\over r}D_{ks}(\omega)K_1(\alpha_sr)\right]\right\} \sin\left({k\pi\over L}z\right) \nonumber \\ 
&=&\left({k\pi\over L}\right)\left\{{\displaystyle\sum_{s=1}^2}A_s\gamma_s\left[C_{ks}(\omega)I_0(\alpha_sr)-{2\mu\over r}D_{ks}(\omega)I_1(\alpha_sr)\right]\right. \nonumber \\
&&~+~\left.{\displaystyle\sum_{s=1}^2}B_s\gamma_s\left[C_{ks}(\omega)K_0(\alpha_sr)+{2\mu\over r}D_{ks}(\omega)K_1(\alpha_sr)\right]\right\} \sin\left({k\pi\over L}z\right) \nonumber \\ 
&=&\left\{{\displaystyle\sum_{s=1}^2}A_s\left[\beta_sI_0(\alpha_sr)-{2\mu\alpha_s\over r}I_1(\alpha_sr)\right]\right. \nonumber \\
&&~+~\left.{\displaystyle\sum_{s=1}^2}B_s\left[\beta_sK_0(\alpha_sr)+{2\mu\alpha_s\over r}K_1(\alpha_sr)\right]\right\} \sin\left({k\pi\over L}z\right), \quad
\end{eqnarray}
where relations (\ref{DDEconst1})-(\ref{DDEconst2}) have been employed in obtaining the first equality in (\ref{DDEsigmaRRreduceF}) and definitions (\ref{defnKBessel}) and (\ref{BFiden}) in obtaining the second and third equalities, respectively. The fourth equality in (\ref{DDEsigmaRRreduceF}) derives from use of (\ref{correspconstsC1}), and the final equality in (\ref{DDEsigmaRRreduceF}) then follows from definitions (\ref{ERPconstsRADSTRSS}) and the fact that 
\begin{eqnarray}\label{gammaequiv}
\gamma_1=1, \quad \gamma_2={1\over\left({k\pi\over L}\right)^2}\left[\left({k\pi\over L}\right)^2-{\rho\omega^2\over\mu}\right]={1\over\left({k\pi\over L}\right)^2}\left\{\begin{array}{rrr}
             \alpha^2_2 & ~~\text{Case~1} \\
             -\alpha^2_2 & ~~\text{Case~2} \\
             -\alpha^2_2 & ~~\text{Case~3} \\
             \end{array} \right..
\end{eqnarray}

For Case 1 and longitudinal wave number $k\neq0$, the axisymmetric shear stress obtained from our method of solution is (again omitting the $\sin(\omega t)$ factor): 
\begin{eqnarray}\label{ourRZstrsscompC1meq0}
\sigma^{(SC)}_{rz}=\mu\left({k\pi\over L}\right)\left\{\sum_{s=1}^2\alpha_s(1+\gamma_s)\Big[A_sI_1(\alpha_sr)-B_sK_1(\alpha_sr)\Big]\right\}\cos\left({k\pi\over L}z\right),  
\end{eqnarray}
which is immediately obtained upon substituting $m=0$ into Eq.~(\ref{strsscmpRZC1}). 

Given the remarks at the beginning of Section \ref{ReducEquivERP}, the ERP shear stress component (\ref{DDEsigmaRZ}) reduces to the following:
\begin{eqnarray}\label{DDEsigmaRZreduce1}
\sigma^{(ERP)}_{rz}&=&-\mu\Omega_2\Big[QJ_1(\Omega_2r)+RY_1(\Omega_2r)\Big] \nonumber \\
&+&\mu\left\{{\displaystyle\sum_{s=1}^2}\left(-k_{rks}+\chi_{ks}k_{zk}\right)\Big[Q_{ks}J_1(k_{rks}r)+R_{ks}Y_1(k_{rks}r)\Big]\right\}\cos(k_{zk}z). \quad
\end{eqnarray}
Employing relations (\ref{DDEconst1})-(\ref{DDEconst4}) and considering only Case 1, the frequency-dependent ERP constants in (\ref{DDEsigmaRZreduce1}) are as follows:
\begin{eqnarray}\label{ERPconstsSHEARSTRSS}
\xi_{ks}(\omega)\equiv-k_{rks}+\chi_{ks}k_{zk}&=&\left\{\begin{array}{lll}
             ~2\alpha_1i^{-1}, & ~~\text{if}~s=1 \\ 
             {\displaystyle\left(\alpha_2+{\left({k\pi\over L}\right)^2\over\alpha_2}\right)}i^{-1}, & ~~\text{if}~s=2 \\
             \end{array} \right. \nonumber \\ \nonumber \\
&=&\left[{\alpha_s(1+\gamma_s)\over\gamma_s}\right]i^{-1}, 
\end{eqnarray}
where the final equality in (\ref{ERPconstsSHEARSTRSS}) derives from use of identity (\ref{gammaequiv}). Then, using the compact notation
\begin{subequations}\label{compactNotSHEARSTRSS} 
\begin{equation}
\bar{\sigma}^{(ERP)}_{rz}\equiv-\mu\Omega_2\Big[QJ_1(\Omega_2r)+RY_1(\Omega_2r)\Big], 
\end{equation}
\begin{equation}
\bar{\xi}_{ks}(\omega)\equiv i\xi_{ks}(\omega)={\alpha_s(1+\gamma_s)\over\gamma_s}, 
\end{equation}
\end{subequations}
and keeping in mind that we are considering Case 1, the ERP shear stress component (\ref{DDEsigmaRZreduce1}) reduces as follows:
\begin{eqnarray}\label{DDEsigmaRZreduceF}
\sigma^{(ERP)}_{rz}&=&\bar{\sigma}^{(ERP)}_{rz}+\mu\left\{{\displaystyle\sum_{s=1}^2}\bar{\xi}_{ks}(\omega)i^{-1}\Big[Q_{ks}J_1(i\alpha_sr)+R_{ks}Y_1(i\alpha_sr)\Big]\right\}\cos\left({k\pi\over L}z\right) \nonumber \\
&=&\bar{\sigma}^{(ERP)}_{rz}+\mu\left\{{\displaystyle\sum_{s=1}^2}\bar{\xi}_{ks}(\omega)\Big[(Q_{ks}+iR_{ks})i^{-1}J_1(i\alpha_sr)+{2\over\pi}R_{ks}K_1(\alpha_sr)\Big]\right\}\cos\left({k\pi\over L}z\right) \nonumber \\ 
&=&\bar{\sigma}^{(ERP)}_{rz}+\mu\left\{{\displaystyle\sum_{s=1}^2}\bar{\xi}_{ks}(\omega)\Big[\widetilde{Q}_{ks}I_1(\alpha_sr)+\widetilde{R}_{ks}K_1(\alpha_sr)\Big]\right\}\cos\left({k\pi\over L}z\right) \nonumber \\ 
&=&\bar{\sigma}^{(ERP)}_{rz}+\mu\left({k\pi\over L}\right)\left\{{\displaystyle\sum_{s=1}^2}\bar{\xi}_{ks}(\omega)\gamma_s\Big[A_sI_1(\alpha_sr)-B_sK_1(\alpha_sr)\Big]\right\}\cos\left({k\pi\over L}z\right) \nonumber \\ 
&=&-\mu\Omega_2\Big[QJ_1(\Omega_2r)+RY_1(\Omega_2r)\Big] \nonumber \\
&+&\mu\left({k\pi\over L}\right)\left\{\sum_{s=1}^2\alpha_s(1+\gamma_s)\Big[A_sI_1(\alpha_sr)-B_sK_1(\alpha_sr)\Big]\right\}\cos\left({k\pi\over L}z\right), 
\end{eqnarray}
where relations (\ref{DDEconst1})-(\ref{DDEconst4}) have been employed in obtaining the first equality in (\ref{DDEsigmaRZreduceF}), and then (\ref{defnKBessel}), (\ref{BFiden}), (\ref{correspconstsC1}), and (\ref{compactNotSHEARSTRSS}), in obtaining the second, third, fourth, and fifth equalities, respectively. 

The first bracketed term in the final line of (\ref{DDEsigmaRZreduceF}) is consistent with the shear stress component (\ref{strsssolnkeq0B}) when $m=0$. Comparing the final line of (\ref{DDEsigmaRRreduceF}) with (\ref{OURsigmaRRC1meq0}) and the cosine term in the final line of (\ref{DDEsigmaRZreduceF}) with (\ref{ourRZstrsscompC1meq0}), we see that, in Case 1, the axisymmetric stress components obtained from the ERP method are identical to the axisymmetric ($m=0$) stress components obtained from our method of solution. Using relations (\ref{DDEconst1})-(\ref{DDEconst4}), definitions (\ref{defnKBessel})-(\ref{BFiden}), identity (\ref{gammaequiv}), and analogs of substitutions (\ref{correspconstsC1}), and applying the same logic, it is straightforward (albeit tedious) to show that the equivalence holds for Cases 2 and 3 as well. Thus, in the special $m=0$ case, the general stress field obtained from our method of solution is identical to the general axisymmetric stress field obtained from the ERP method.  

\begin{acknowledgments}
The authors acknowledge financial support from the Natural Sciences and Engineering Research Council (NSERC) of Canada and the Ontario Research Foundation (ORF). 
\end{acknowledgments}


\begin{thebibliography}{99}

%\bibitem{Glad75} G.~M.~L. Gladwell, D.~K. Vijay, Natural frequencies of free finite-length circular cylinders, J. Sound Vib. 42 (1975) 387-397.

\bibitem{Soldatos94} K.P. Soldatos, Review of three dimensional dynamic analyses of circular cylinders and cylindrical shells, Appl. Mech. Rev. 47 (1994) 501-516. 

\bibitem{Qatu02} M.S. Qatu, Recent research advances in the dynamic behavior of shells: 1989-2000, Part 2: Homogeneous shells, Appl. Mech. Rev. 55 (2002) 415-434.

\bibitem{VTCS10} H.R. Hamidzadeh, R.N. Jazar, Vibrations of Thick Cylindrical Structures, Springer, New York, 2010.  

\bibitem{LeissaQatu11} A.W. Leissa, M.S. Qatu, Vibrations of Continuous Systems, McGraw-Hill, New York, 2011. 

\bibitem{Hutch86} J.~R. Hutchinson, S.~A. El-Azhari, Vibrations of free hollow circular cylinders, J. Appl. Mech. 53 (1986) 641-646. 

\bibitem{Soldatos90} K.~P. Soldatos, V.~P. Hadjigeorgiou, Three dimensional solution of the free vibration problem of homogeneous isotropic cylindrical shells and panels, J. Sound Vib. 137 (1990) 369-384. 

\bibitem{Leissa97} J. So, A.~W. Leissa, Free vibrations of thick hollow circular cylinders from three-dimensional analysis, J. Vib. Acoust. 119 (1997) 89-95. 

\bibitem{Loy99} C.T. Loy, K.Y. Lam, Vibration of thick cylindrical shells on the basis of three-dimensional theory of elasticity, J. Sound Vib. 226 (1999) 719-737.

\bibitem{Buchanan02} G.~R. Buchanan, C.~B.~Y. Yii, Effect of symmetrical boundary conditions on the vibration of thick hollow cylinders, Appl. Acoust. 63 (2002) 547-566. 

\bibitem{Zhou03} D. Zhou, Y.K. Cheung, S.H. Lo, F.T.K. Au, 3D vibration analysis of solid and hollow circular cylinders via Chebyshev-Ritz method, Comput. Method. Appl. M. 192 (2003) 1575-1589. 

\bibitem{Mofak06} M.~R. Mofakhami, H.~H. Toudeshky, Sh.~H. Hashemi, Finite cylinder vibrations with different end boundary conditions, J. Sound Vib. 297 (2006) 293-314. 

\bibitem{MFZK08} P. Malekzadeh, M. Farid, P. Zahedinejad, G. Karami, Three-dimensional free vibration analysis of thick cylindrical shells resting on two-parameter elastic supports, J. Sound Vib. 313 (2008) 655-675.

\bibitem{Khalili12} S.M.R. Khalili, A. Davar, K.M. Fard, Free vibration analysis of homogeneous isotropic circular cylindrical shells based on a new three-dimensional refined higher-order theory, Int. J. Mech. Sci. 56 (2012) 1-25.

\bibitem{Ye14} T. Ye, G. Jin, S. Shi, X. Ma, Three-dimensional free vibration analysis of thick cylindrical shells with general end conditions and resting on elastic foundations, Int. J. Mech. Sci. 84 (2014) 120-137.

\bibitem{Ebenezer15} D.D. Ebenezer, K. Ravichandran, C. Padmanabhan, Free and forced vibrations of hollow elastic cylinders of finite length, J. Acoust. Soc. Am. 137 (2015) 2927-2938. 

\bibitem{Wein74} L.~I. Weingarten, H. Reismann, Forced motion of cylindrical shells, Z. Angew. Math. Mech. 54 (1974) 181-191. 

\bibitem{PJ60s} C. Prasad, R.~K. Jain, Vibrations of transversely isotropic cylindrical shells of finite length, J. Acoust. Soc. Am. 38 (1965) 1006-1009. 
 
\bibitem{Arm69} A.E. Armen\`{a}kas, D.C. Gazis, G. Herrmann, Free Vibrations of Circular Cylindrical Shells, Pergamon Press, Oxford, 1969. 

%\bibitem{MW06} C.~K. Mechefske, F. Wang, Theoretical, numerical, and experimental modal analysis of a single-winding gradient coil insert cylinder, Magn. Reson. Mater. Phy. 19 (2006) 152-166. 

\bibitem{usB1} J. Sakhr, B.~A. Chronik, Solving the Navier-Lame equation in cylindrical coordinates using the Buchwald representation: Some parametric solutions with applications, Adv. Appl. Math.  Mech. 10 (2018) 1025-1056. 

\bibitem{Elastobuch} A.~C. Eringen, E.~S. Suhubi, Elastodynamics, Volume 2, Linear Theory, Academic Press, New York, 1975. 

\bibitem{Wills96} H. Wang, K. Williams, Vibrational modes of thick cylinders of finite length, J. Sound Vib. 191 (1996) 955-971. 

\bibitem{Wills98} H. Wang, K. Williams, W. Guan, A vibrational mode analysis of free finite-length thick cylinders using the finite element method, J. Vib. Acoust. 120 (1998) 371-377.

\bibitem{Wills02} R.~K. Singhal, W. Guan, K. Williams, Modal analysis of a thick-walled circular cylinder, Mech. Syst. Signal Pr. 16 (2002) 141-153. 
 
%\bibitem{Chau94} K.~T. Chau, Vibrations of transversely isotropic finite circular cylinders, J. Appl. Mech. 61 (1994) 964-970. 

%\bibitem{Honarvar09} F. Honarvar, E. Enjilela, A.N. Sinclair, Asymmetric and axisymmetric vibrations of finite transversely isotropic circular cylinders, Acoust. Phys. 55 (6) (2009) 708-714. 

%\bibitem{Liu12} S.~X. Liu, L.~G. Tang, X.~M. Xu, Transient elastodynamic response of finite and infinite solid cylinders, Appl. Acoust. 73 (2012) 798-802. 

%\bibitem{Buchanan01} C.~B.~Y. Yii, G.~R. Buchanan, The effect of the Poisson ratio on the vibration of hollow circular finite-length cylinders, J. Sound Vib. 248 (2001) 187-194. 

\end{thebibliography}
\end{document}